\documentclass[%
reprint,
amsmath,amssymb, nofootinbib,
aps, groupedaddress,superscriptaddress, prl
]{revtex4-2}

\usepackage{graphicx}% Include figure files
\usepackage{dcolumn}% Align table columns on decimal point
\usepackage{bm}% bold math
\usepackage{braket}
\usepackage{mathtools}
\usepackage{bbm}
\usepackage{comment}
\usepackage{makecell}
\usepackage{afterpage}
\usepackage{physics}
\usepackage[usenames,dvipsnames]{xcolor}

\usepackage{soul}

\newcommand\blankpage{%
    \null
    \thispagestyle{empty}%
    \addtocounter{page}{-1}%
    \newpage}

\usepackage{hyperref}
\hypersetup{
	colorlinks=true,
	linkcolor=blue,
	citecolor=blue,
	filecolor=magenta,
	urlcolor=cyan
}

\def\hL{\mathcal{L}}

\graphicspath{{../}}
\def\umd{Joint Quantum Institute, NIST/University of Maryland, College Park, Maryland 20742, USA}
\def\qcs{Joint Center for Quantum Information and Computer Science, NIST/University of Maryland, College Park, Maryland 20742 USA}
\newcommand{\TUM}{\affiliation{Department of Physics, Technical University of Munich, 85748 Garching, Germany}}
\newcommand{\MCQST}{\affiliation{Munich Center for Quantum Science and Technology (MCQST), Schellingstr. 4, 80799 M{\"u}nchen, Germany}}
\begin{document}

\def\papertitle{{Candidate for a passively protected quantum memory in two dimensions}}
\title{\papertitle}

\author{Simon Lieu$^\dagger$}
\affiliation{\umd}
\affiliation{\qcs}

\author{Yu-Jie Liu$^\dagger$} 
\TUM \MCQST
\def\thefootnote{$\dagger$}\footnotetext{These authors contributed equally.}

\author{Alexey V.~Gorshkov}
\affiliation{\umd}
\affiliation{\qcs}

\date{\today}

\begin{abstract}

An interesting problem in the field of quantum error correction involves finding a physical system that hosts a ``passively protected quantum memory,'' defined  as an encoded qubit  coupled to an environment that naturally wants to correct errors.  To date, a quantum memory stable against finite-temperature effects is only known in four spatial dimensions or higher. Here, we take a different approach to realize a  stable  quantum memory by relying on a driven-dissipative environment. We propose a new model, the photonic-Ising model, which  appears to passively correct against both bit-flip and phase-flip errors in two dimensions: A square lattice composed of photonic ``cat qubits'' coupled  via dissipative terms which  tend to fix errors locally. Inspired by the presence of two distinct $\mathbb{Z}_2$-symmetry-broken phases, our scheme relies on Ising-like dissipators to protect against  bit flips and  on a driven-dissipative photonic environment to protect against phase flips.  We also discuss possible ways to realize the photonic-Ising model.

\end{abstract}

\maketitle

Quantum error correction remains one of the biggest challenges towards building a practical quantum computer \cite{nielsen_book, lidar_book}. 
One of the leading candidates for realizing fault tolerance is the family of quantum stabilizer codes~\cite{gottes1997}, including the surface code~\cite{bravyi:1998,fowler:2012,eric2002} and the GKP code~\cite{gottesman:2001}. These error-correcting schemes are based on fast error recovery controlled by the feedback from repetitive syndrome measurements.

A prominent alternative is the finite-temperature quantum memory:
Certain thermal environments naturally evolve arbitrary initial states into a qubit subspace of interest at low temperature, thus eliminating the need for active measurements and correcting operations. Many recent studies have investigated thermal self-correcting properties \cite{bacon2006, yoshida2011, roberts2020b,terhal2015, brown2016, bombin2013, haah2011, bravyi2013, loss2010, breuckmann2016, Alicki2009, Alicki2010, eric2002}.
 To date, the only known models that host a passive quantum memory via this mechanism are topological codes in four dimensions (4D) and higher, e.g.~the 4D toric code \cite{Alicki2010, eric2002}. 

A separate line of research aims to uncover a passively protected quantum memory via engineered ``driven-dissipative’’ systems \cite{passive1,passive2,passive3,passive4,passive5, passive6, passive7,  pastawski2011, mirrahimi2014, kapit2016, reiter2017, fujii2014, cat-exp3,gyenis2021,CampagneIbarcq2020,cat-exp1,cat-exp2, lescanne2020, berdou2022, lieu2020}. Such passive protection includes but is not limited to the finite-temperature case, since a thermal-equilibrium steady state is not required. The memory is dynamically protected against certain noise channels by (local) Markovian dissipation. 
 This has led to a number of new ideas for passive error correction, such as
the autonomously corrected cat qubit 
%via the fast reset of a cat qubit
~\cite{chen2021,leghtas2013}
%the $0$-$\pi$ qubit~\cite{benoit:2002,ioffe:2002,brooks:2013}, or 
and the dissipative  Toom's rule~\cite{toom,pastawski2011,eric2002}. Unfortunately, none of these models can protect a quantum memory for an exponentially-long time as a function of the system size (in less than four dimensions).

%models can fix both bit flips and phase flips without the presence of measurements. 

%, the proposed model provides an example of a robust quantum memory which, at low temperature, can be exponentially long-lived in %some 
%system size parameters and has challenging, yet realistic
%amenable 
%physical requirements. 

In this work, we study a model with engineered dissipation which appears to protect against both bit flips and phase flips and lives in two spatial dimensions. Instead of relying on topological order, we suggest  that the model should  belong to a phase that spontaneously breaks two different $\mathbb{Z}_2$ symmetries.  Each $\mathbb{Z}_2$-symmetry-broken phase protects a ``classical bit,'' which together form a robust qubit.  The proposed model provides an example of a robust quantum memory which, at low temperature, can be exponentially long-lived in %some 
system size parameters and has challenging, yet realistic
%amenable 
physical requirements.

%\change{Compared to previously known schemes of passive quantum error correction, such as the autonomously corrected cat qubit 
%via the fast reset of a cat qubit
%~\cite{chen2021,leghtas2013}, the $0$-$\pi$ qubit~\cite{benoit:2002,ioffe:2002,brooks:2013}, or the dissipative 4D toric code~\cite{pastawski2011,eric2002}, the proposed model provides an example of a robust quantum memory which, at low temperature, can be exponentially long-lived in %some 
%system size parameters and has challenging, yet realistic
%amenable 
%physical requirements. 

\textit{Quantum memory.}---Consider a Hilbert space $\mathcal{H}$, and define two encoded, logical states $| \bar 0 \rangle,| \bar 1 \rangle \in\mathcal{H}$ that span the codespace $\mathcal{C}$. We assume the system is always initilized in the codespace: $\rho_i = | \psi \rangle \langle \psi |$ where $\ket{\psi}\in\mathcal{C}$.

A local continuous-time Markovian generator $\mathcal{L}$ in Lindblad form is defined by
\begin{align} \label{eq:lindblad}
    \frac{d \rho}{dt}  = \hL(\rho) = -i[H,\rho] +\sum_j  \left(L_j\rho L_j^{\dag}-\frac{1}{2}\{L^{\dag}_j L_j,\rho\}\right),
\end{align}
where $H$ is the Hamiltonian of the system and $L_j$ are local dissipators which arise due to the system-environment coupling \cite{lindblad1976}. We  consider a dynamical process that can be decomposed into two parts,  an ``error'' generator and a ``recovery'' generator: $\hL =  \hL_e + \hL_r $. The error generator describes the main channels of physical noise which move the initial state out of the codespace. The recovery generator stabilizes the codespace: $\hL_r(\rho_i) = 0$, i.e.~any state in the codespace is a steady state of the recovery.  We allow for this noisy process to occur for a time $t$, which generically sends $\rho_i$ to a mixed state $\rho_m(t) = e^{\mathcal{L} t}  (\rho_i)$.

Finally, we employ a ``single-shot'' decoding quantum channel $\mathcal{E}_r$ which sends every state in the Hilbert space back to the codespace  \cite{fn1}. The final state is
\begin{equation}
\rho_f(t) = \mathcal{E}_r e^{\mathcal{L} t} (\rho_i).
\end{equation}

We wish to  find systems  where the difference between the initial and final states is exponentially small  in the system size:
\begin{align} \label{eq:exp_small}
1- \Tr[\rho_i \rho_f(t)]  =O( e^{-\gamma M})\text{ as }M\to\infty,
\end{align}
where $\gamma> 0$ is a time-independent constant and $M$ is  some system size parameter. A system described by $\hL$ hosts a passively protected quantum memory for any finite time $t$ if Eq.~\eqref{eq:exp_small} holds as the thermodynamic limit is approached.
% The recovery Lindbladian $\hL_r$ ensures that  errors do not corrupt quantum information for any finite $t$.

The bit-flip and phase-flip errors of a two-level system are generated via the  Pauli operators $X, Z $ respectively. A good quantum memory should thus protect against both sources of noise. Recent work \cite{lieu2020} has described the connection between $\mathbb{Z}_2$ symmetry breaking  and error correction: A symmetry-broken phase  protects quantum information against $X$ \textit{or} $Z$ errors, but not both. This leads to a protected classical bit, which can be viewed as a quantum bit experiencing biased noise \cite{amazon_cat}. 

In this work,  we attempt to glue two different  classical bits together to form a robust qubit. Our strategy involves studying a system that  passively corrects against  bit flips due to Ising-like dissipators which tend to align qubits locally. Furthermore, phase flips will passively correct due to driven-dissipative stabilization of the photonic cat code.  
We begin by describing spontaneous symmetry breaking  in the cat code and in the Ising model separately. We then  describe a model which  inherits both protecting features.

\textit{Photonic cat code.}---Let us briefly review $\mathbb{Z}_2$ spontaneous symmetry breaking in the
%a single 
photonic cat code \cite{mirrahimi2014, knight1994}. [For a detailed analysis, we refer to Ref.~\cite{lieu2020}.] Consider a driven-dissipative photonic cavity  in the presence of two-photon drive and two-photon loss. The rotating-frame Hamiltonian  and dissipator read
$
H = \lambda \left(a^2 + (a^\dagger) ^2 \right), \  L_2 = \sqrt{\kappa_2} a^2.
$
Here $a$ is the annihilation operator for a cavity photon, $\lambda$ is the drive strength, and $\kappa_2$ is the two-photon loss rate.
While the model has  $\mathbb{Z}_2$ symmetry $[H,Q] = [L_2, Q] = 0$ generated by parity $Q = e^{i \pi a^\dagger a}$, the steady state can violate this symmetry:
\begin{align}
\rho_{ss} = |\psi \rangle \langle \psi |, \qquad  |\psi \rangle  = c_0 |\alpha_e \rangle +  c_1 |\alpha_o \rangle, 
\end{align}
for $|c_0|^2 + |c_1|^2 =1$, where $|\alpha_e \rangle \sim |\alpha \rangle + | - \alpha \rangle, |\alpha_o \rangle \sim |\alpha \rangle - | - \alpha \rangle $, and $|\alpha \rangle$ is a coherent state with amplitude $\alpha = e^{-i \pi / 4} \sqrt{N}$ and $N \equiv \lambda / \kappa_2$ photons. The even and odd cat states $|\alpha_{e/o} \rangle$ represent logical 0 and 1, respectively.

The cat code
is protected against phase-flip errors generated by photon dephasing $L_d = \sqrt{\kappa_d} a^\dagger a$. Indeed, the phase-flip logical error rate scales as $e^{-\gamma N}$ where $\gamma$ is a constant~\cite{mirrahimi2014}. 
The symmetry-broken states $|\pm \alpha \rangle \approx  (|\alpha_e \rangle \pm |\alpha_o \rangle) /\sqrt{2} $ have an exponentially-long lifetime in the limit of large $N$, ensuring that logical phase flips are unlikely.

The dominant decoherence mechanism for the cat qubit stems from the  bit flip, generated via single-photon loss $L_1 = \sqrt{\kappa_1} a$: $a |\alpha_{e/o} \rangle \sim |\alpha_{o/e} \rangle   $, which reduces the qubit steady state structure to a classical bit: $\rho_{ss} \approx c |+\alpha \rangle \langle +\alpha | + (1-c) |-\alpha \rangle \langle -\alpha |, c\in[0,1] $~\cite{lieu2020}.  More generally, perturbations that commute with photon parity (e.g.~$[L_d, Q]=0$) are expected to be passively corrected, while terms which explicitly break the symmetry (e.g.~$\{ L_1, Q \}=0$) are not.

\textit{2D Ising model.}---We now turn our attention to a system that has the opposite problem: $\mathbb{Z}_2$ symmetry breaking will protect against bit flips but not phase flips. 
%
% We design a local Lindbladian  such that its steady state is the thermal state of the   2D Ising model on an $M \times M$ lattice with periodic boundary conditions. The  2D   Ising model Hamiltonian reads
We consider the 2D Ising model on an $M \times M$ square lattice with periodic boundary conditions. The Hamiltonian reads
 \begin{align}
 H_{is} = - \sum_{x,y = 1}^M  (Z_{x,y} Z_{x+1,y} + Z_{x,y} Z_{x,y+1}) \,,
 \end{align}
 where $Z_{x,y}$ is the $Z$ Pauli operator on site $(x,y)$. The ferromagnetic states are the ground states of this model and span the codespace: $|\bar 0\rangle \equiv| \downarrow  \downarrow  \downarrow  \ldots \rangle , |\bar 1\rangle \equiv | \uparrow  \uparrow  \uparrow  \ldots \rangle$, with $Z\ket{\downarrow}= \ket{\downarrow}$ and $Z\ket{\uparrow}= -\ket{\uparrow}$.

% We define dissipators  and bit flips that locally obey detailed balance with respect to this Hamiltonian. 
We define local dissipators that describe the thermalization of the Ising Hamiltonian.
(For simplicity, we set the  Hamiltonian in the master equation to zero.) Consider dissipators  that are a product of a spin flip ($X$) with a projector onto a particular domain-wall configuration. These jumps will cause a spin  to flip sign according to a local ``majority rule,'' i.e.~only if  more than two of the neighboring spins are misaligned. Specifically:
\begin{align}
L_{x,y}^{(4)} &= \sqrt{\kappa} X_{x,y} P_{x,y;\rightarrow}^- P_{x,y;\uparrow}^- P_{x-1,y;\rightarrow}^- P_{x,y-1;\uparrow}^- \,,
\nonumber\\
L_{x,y}^{(3)} &=  \sqrt{ \tilde{\kappa}}  X_{x,y} P_{x,y;\rightarrow}^+ P_{x,y;\uparrow}^- P_{x-1,y;\rightarrow}^- P_{x,y-1;\uparrow}^- \,,  \label{eq:Ising_nn}
\end{align}
where $\tilde{\kappa} = \sqrt{ \Delta \kappa + \Delta^2} -\Delta $ and  $P_{x,y;\rightarrow}^{\pm} = (1 \pm Z_{x,y}  Z_{x+1,y})/2$ , $P_{x,y;\uparrow}^{\pm} = (1 \pm Z_{x,y}  Z_{x,y+1})/2  $ are projectors onto particular local configurations of spins. 
The superscripts indicate the number of domain walls  which  the projector is checking for, and we neglect to write jumps related by rotational invariance (i.e.~there are 4 different $L^{(3)} $ operators per site) \cite{fn3}.  We also  consider an error process in the form of a uniform bit flip rate on each lattice site:
$
L_{x,y}' = \sqrt{\Delta} X_{x,y}.
$.

\begin{figure}
    \centering
    \includegraphics[scale=0.4]{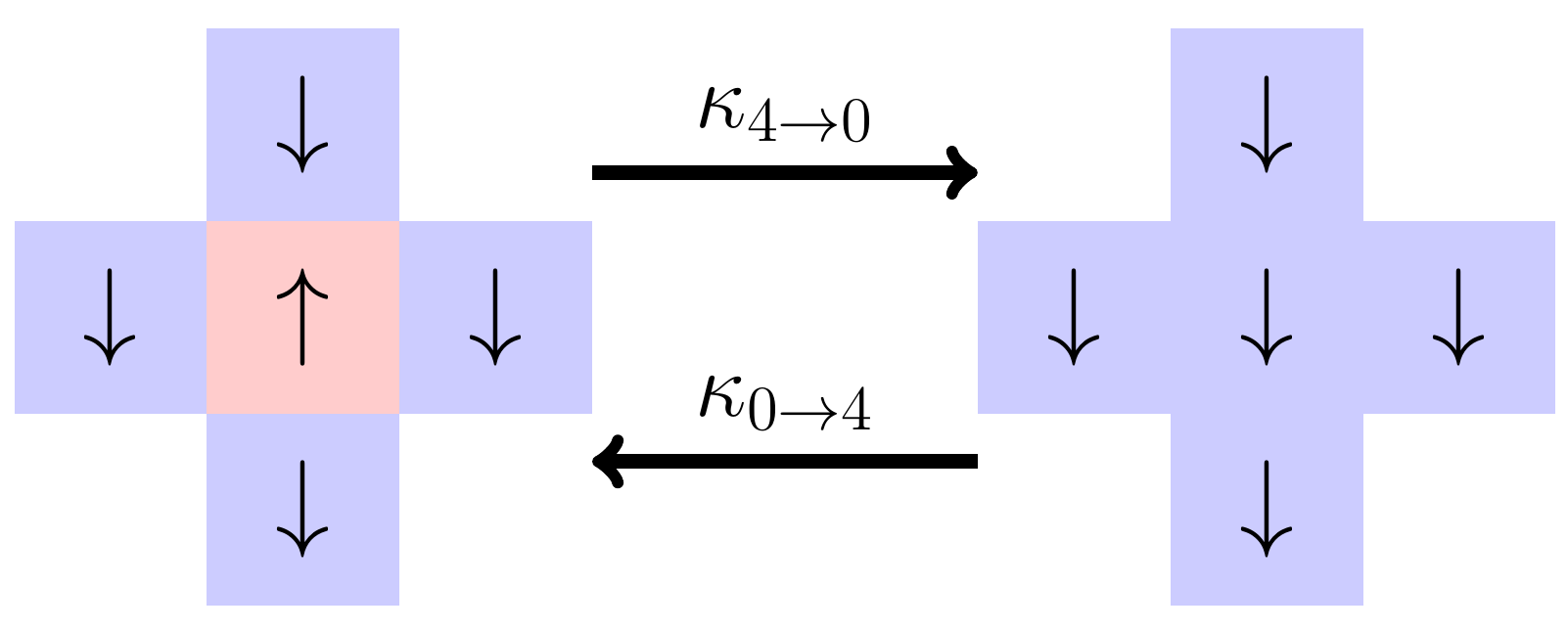}
    \caption{ The total rate of transitioning from a configuration with 4 domain walls to a configuration with 0 domain walls satisfies detailed balance: $\kappa_{4 \rightarrow 0 } / \kappa_{0 \rightarrow 4}  = e^{8 \beta}$.} 
    
    \label{fig:ising_rates}
\end{figure}

We have chosen our dissipators above such that the steady state of the model is the thermal state of the 2D classical Ising model:
\begin{align} \label{eq:thermal_ss}
\rho_{ss}  =  \frac{e^{- \beta H_{is}}}{\Tr[e^{- \beta H_{is}}]}, \qquad \beta = \frac{1}{8} \ln \left[\frac{\kappa + \Delta}{\Delta} \right],
\end{align}
with the effective temperature  set by the relative ratio of the correction rate to the bit-flip rate.   Within the quantum jump picture~\cite{knight1998, daley2014}, the rates of transitioning between different classical configurations respect detailed balance. (See e.g.~Fig.~\ref{fig:ising_rates}.)

While the thermal state \eqref{eq:thermal_ss} is always a steady state of the model,  it is not  unique.  All dissipators commute with the parity operator $Q =\prod_{i=1}^{M^2} X_i $:  $[L_j, Q] = 0$. This means that the dynamics preserves the parity of the state (called a ``strong $\mathbb{Z}_2$ symmetry'' \cite{prosen2012}).  In the thermodynamic limit of the low-temperature (symmetry-broken) phase, a qubit can be stored in the steady state \cite{lieu2020}.

\begin{figure}
    \centering
    \includegraphics[scale=0.28]{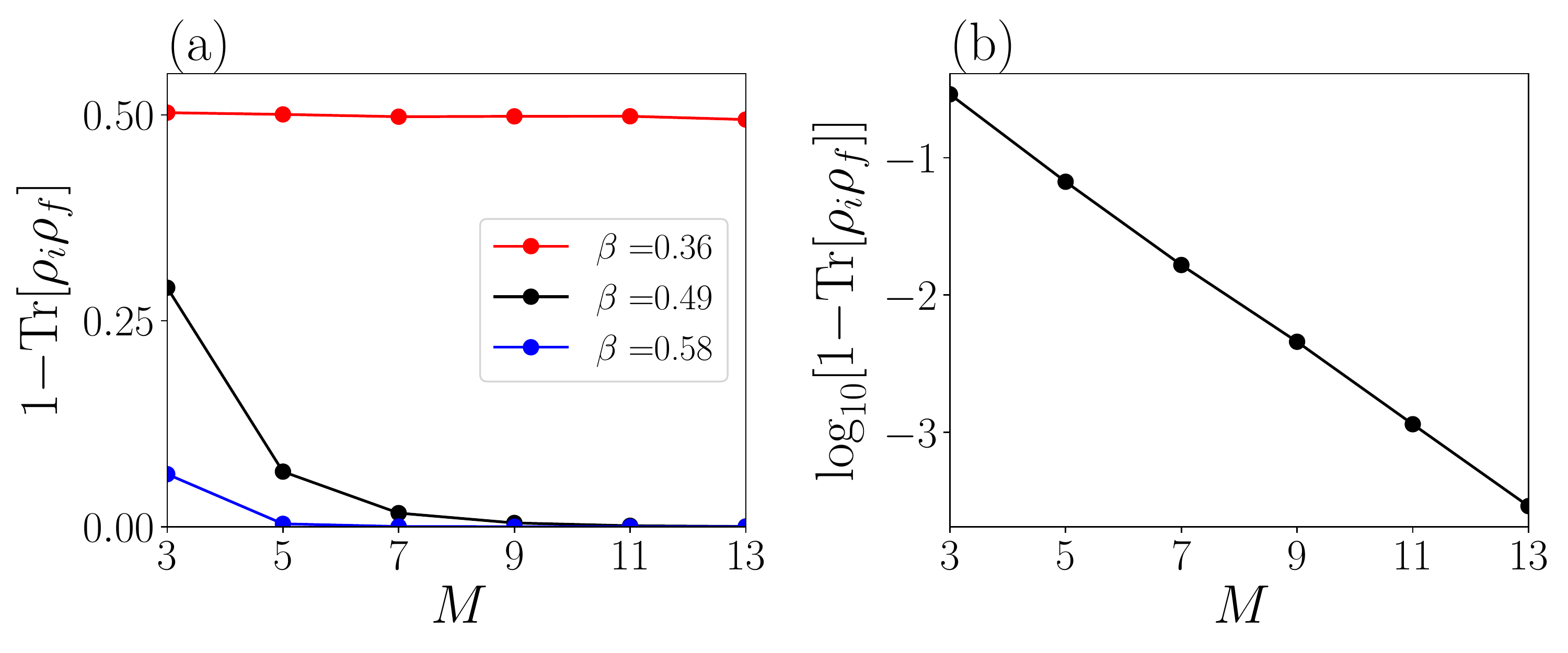} 
    
    \caption{  (a) The overlap between the initial and final states for the protocol given in the main text, for a Lindbladian in the high-temperature phase (red dots), and in the low-temperature phase (black and blue dots).  As linear system size $M$ grows, the overlap approaches one  only in the low-temperature (symmetry-broken) phase corresponding to $\beta > \beta_c \approx 0.44$. (b) Same black data points on a log plot; the overlap tends to one exponentially fast in $M$. In both (a) and (b), 
    %we use $\kappa = 1$ and 
    the quench time is $T = 800 / \kappa $, i.e.~long enough to reach the steady state. The simulation employs the quantum jump approach by averaging over $10^5$ trajectories.}
    \label{fig:ising_error_correction}
\end{figure}

We can confirm this picture via numerical simulations.  
Suppose we initialize our system in a ferromagnetic state: $| \psi \rangle = | \bar{0} \rangle = (| E_0^+ \rangle + | E_0^- \rangle) /\sqrt{2} $ where $| E_0^\pm \rangle$ are ground states in the different parity sectors \cite{fn2}. We then quench the system with the noisy Lindbladian for a time $T$ much larger than the inverse of the dissipative gap, so that the system settles into its steady state. Finally, we apply a single-shot decoder which brings the state back to the codespace  by measuring all domain walls in the system then flipping all bits in the smaller domain. Our results are summarized in Fig.~\ref{fig:ising_error_correction}. In the low-temperature phase, the overlap starts to approach the ideal value of 1 exponentially fast in $M$.  Qualitatively different behavior occurs in the high-temperature phase [$\beta > \beta_c = \ln(1+\sqrt{2}) / 2\approx 0.44$;  red dots], where the success rate stays at 50\% for a wide range of $M$.

%Since domain walls cost an energy proportional to their perimeter, it is exponentially unlikely to flip a macroscopic number of spins in the thermodynamic limit of the low-temperature (symmetry-broken) phase, $\beta > \beta_c = \ln(1+\sqrt{2}) / 2\approx 0.44$. 

Unfortunately, the stored qubit is unstable to noise that violates the strong symmetry. In particular, the presence of dephasing $L_i \sim Z_i$ (phase flips), reduces the strong $\mathbb{Z}_2$ symmetry to a ``weak $\mathbb{Z}_2$ symmetry'' (defined at the level of the superoperator: $[\hL, \mathcal{Q}]=0$, where  $\mathcal{Q}(\rho) = Q \rho Q^{\dag}$), such that only a classical bit can be stored in the steady state. In this case, the steady state at low-temperature has the structure $\rho_{ss}  \approx c  |\bar{0}\rangle \langle \bar{0}| + (1-c)|\bar{1} \rangle  \langle \bar{1}|$,
for $c \in [0,1]$. In analogy with the  cat qubit  in the presence of single-photon loss, $Z$ dephasing destroys the coherence between Ising ferromagnetic states.

\textit{2D photonic-Ising model.---}We see that the cat code  passively corrects against phase flips but not bit flips, and that the 2D Ising model passively corrects against bit flips but not phase flips. Is it possible to combine the protecting features of both models to construct a system that passively corrects against both sources of noise?

Consider an $M \times M$ square lattice of photonic cavities. Each cavity undergoes a  two-photon drive  process and a two-photon loss process:
\begin{align}\label{eq:lattice_2ph}
    H_{x,y} = \lambda (a_{x,y}^2 + (a^\dagger_{x,y})^2), \qquad L_{2,x,y} = \sqrt{\kappa_2} a_{x,y}^2 \,,
\end{align}
where $a_{x,y}$ is the annihilation operator on site $(x,y)$.  Next, let us consider a parity-parity interaction between neighboring cavities: $H_S = - \sum_{\langle i j \rangle} Q_i Q_j$, where $Q_{j} $ is the photon parity operator at site $j$. Similar to the Ising model, at low temperatures, such interaction will tend to align the parities of neighboring cavities via the following local dissipators~(for a microscopic derivation of the dissipators, see the SM \cite{SM}):
%Next, we consider 
%local nearest-neighbor dissipators of the following form:
\begin{align}
L_{x,y}^{(4)} &= \sqrt{\kappa_{nn}} a_{x,y} P_{x,y; \rightarrow}^- P_{x,y;\uparrow}^- P_{x-1,y;\rightarrow}^- P_{x,y-1;\uparrow}^-\,, \nonumber
\\
L_{x,y}^{(3)} &=  \sqrt{ \tilde{\kappa}_{nn} }  a_{x,y} P_{x,y;\rightarrow}^+ P_{x,y;\uparrow}^- P_{x-1,y;\rightarrow}^- P_{x,y-1;\uparrow}^-\,, \label{eq:nn} 
\end{align}
where $a_{x,y}$ is  the annihilation operator for the cavity at site $x,y$, $\tilde{\kappa}_{nn} = \sqrt{ \kappa_{1}\kappa_{nn} + \kappa_{1}^2} -\kappa_{1} $, $\kappa_1$ is the single-photon loss rate (corresponding to the dissipator: $L_{1,x,y} = \sqrt{\kappa_1} a_{x,y})$, $P^\pm_{x,y;\rightarrow} = (1 \pm Q_{x,y} Q_{x+1,y})/2, P^\pm_{x,y;\uparrow} = (1 \pm Q_{x,y} Q_{x,y+1})/2, $ and $Q_{x,y} = e^{i \pi a_{x,y}^\dagger a_{x,y}} $. The following states are the steady states of the model in the absence of errors ($\kappa_1= 0$) and  span the codespace:
\begin{align}
 |\psi \rangle  = c_0 |\alpha_e \rangle   |\alpha_e \rangle   |\alpha_e \rangle \ldots +  c_1 |\alpha_o \rangle   |\alpha_o \rangle   |\alpha_o \rangle\ldots\,,
\end{align}
for $|c_0|^2 + |c_1|^2 =1$.

For thermal systems, the existence of a passively-correcting quantum memory is related to the presence of an extensive energy barrier  which local errors must overcome in order to create a logical bit-flip or phase-flip operation \cite{bravyi2009}. In the model described above,  a logical bit-flip operation can be created via local single-photon loss $L_{1,x,y} = \sqrt{\kappa_1} a_{x,y}$ only by passing through a configuration with an extensive number of domain walls, which is exponentially unlikely in the limit of large lattice size  $M \rightarrow \infty$. 
Similarly, a phase-flip error can only  be generated by taking the state $|\alpha_e\rangle  \pm |\alpha_o\rangle  $ to $ |\alpha_e\rangle  \mp |\alpha_o\rangle $ for any of the cavities. However,  such a process is also unlikely to occur via  dephasing perturbations $L_{d, x,y}  = \sqrt{\kappa_d} a^\dagger_{x,y} a_{x,y} $  which perturb   states locally in phase space, since  the states $|\pm \alpha\rangle \approx |\alpha_e\rangle  \pm |\alpha_o\rangle$ are well separated in phase space and  an unstable fixed point sits between them  \cite{prx2019}. The logical phase-flip errors are again exponentially unlikely  as $N \rightarrow \infty$.

The single-photon loss and the dephasing lead to terms proportional to $ a^{\dag}a$ and  $ (a^{\dag}a)^2$ in the Lindbladian, which result in leakage out of the effective two-level codespace for each cavity into other states of the cavity. 
This leakage poses a challenge for numerical simulation since (unlike the Ising model) we need to keep track of more than two degrees of freedom per lattice site. Nevertheless, we shall provide evidence for a stable quantum memory  by employing  a variety of approximations. 

First, let us consider an approximation that allows us to map the dynamics of the photonic-Ising model directly to the  classical-Ising model studied above. Specifically, 
we introduce an idealized model by replacing the
single-photon loss dissipator $L_1 = \sqrt{\kappa_1} a$ with $E_1 = \sqrt{\kappa_1} b$, where $b = a V $ and $V$ is the projector onto the codespace: $V = | \alpha_e \rangle \langle \alpha_e |+ | \alpha_o \rangle \langle \alpha_o| $. We also assume an absence of dephasing errors, i.e.~$\kappa_d = 0$. This allows us to treat each site as an effective two-level system $|  0\rangle = | \alpha_e \rangle, 
|  1\rangle = | \alpha_o \rangle$, avoiding any leakage out of the codespace. 
We similarly replace $a \rightarrow b$ in the nearest-neighbor coupling dissipators  \eqref{eq:nn} (except in the definition of $Q$). The operator $b$ can be regarded as an ``idealized bit flip'' since, for $N \gg 1$, it takes the form $b \approx \alpha (|\alpha_e \rangle \langle \alpha_o | + |\alpha_o \rangle \langle \alpha_e |)$.
The idealized model maps exactly to the  Ising model studied above, with an effective bit-flip error rate of $N \kappa_1$, an effective Ising-correction rate of $N \kappa_{nn}$, and an inverse temperature $\beta = \ln \left[ (\kappa_{nn} +  \kappa_1) / \kappa_1 \right]/8.$ We therefore find that  this model passively corrects against bit flips in the limit $M\rightarrow \infty$ of the low-temperature phase. 
In the limit of  large driving strength and small single-photon loss,
we expect the photonic-Ising model to be well approximated by the idealized model since the state rarely leaves the codespace. 
We provide quantitative evidence for this in the Supplemental Material (SM) \cite{SM}.

Dephasing, single-photon loss, and bit-flip recovery jumps ($L^{(3)}_{x,y}$ and  $L^{(4)}_{x,y}$) cause
leakage out of the codespace  which is neglected within the idealized model. It is natural to ask whether 
this leakage  is detrimental to the passively protected memory when the idealized model is no longer a good approximation. We  provide evidence that  this is not the case by  studying a toy model which resembles the 2D model. Consider a  single cavity coupled to a spin-$1/2$ particle (described by Pauli operators $X, Y, Z$), leading to two logical states $\ket{\downarrow}\ket{\alpha_e}$ and $\ket{\downarrow}\ket{\alpha_o}$. The  Hamiltonian and jump operators read $h = \lambda (a^2 + (a^\dagger)^2),\ l_2 = \sqrt{\kappa_2} a^2 ,\ 
 l_1  = \sqrt{\kappa_1} X a,\  l_d = \sqrt{\kappa_d} a^{\dag}a,\
 l_{nn} = \sqrt{\kappa_{nn}} \frac{1}{2}X(1-Z) a$.
The model assumes that single-photon loss is accompanied by a spin flip, while two-photon drive and  dephasing 
are not. The flip-recovery jump $l_{nn}$ is triggered by a flipped spin state $\ket{\uparrow}$, similar to the bit-flip recovery jump caused by a parity misalignment in 2D. Importantly, leakage caused by the noise processes $l_1, l_d$, and the flip-recovery jump is captured by this
model. In the SM \cite{SM}, we analyze this model numerically and analytically. We find that
the initial state can always be  perfectly restored via a decoder (up to corrections exponentially small in $N$).

\begin{figure}
    \centering
    \includegraphics[scale=0.9]{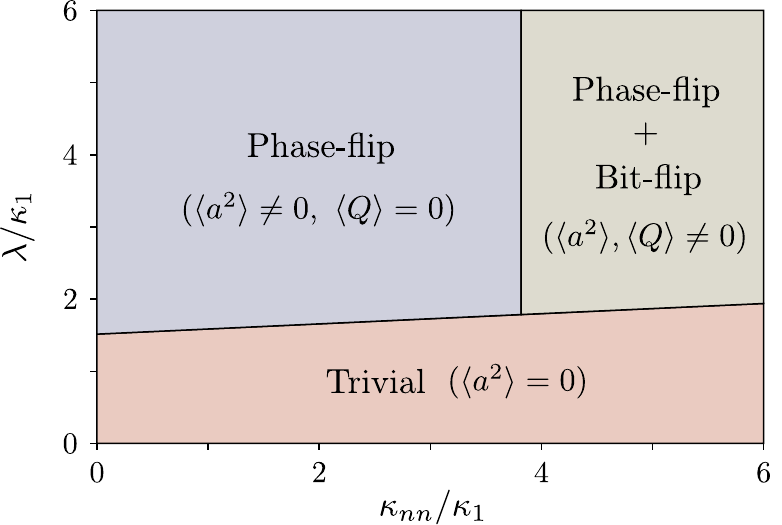}
    \caption{The mean-field phase diagram  for  $\kappa_d = \kappa_1$. The top right corner shades the region where both $\langle Q\rangle$ and $\langle a^2\rangle$ are non-zero. Both phase and bit-flip errors are protected. When $\langle a^2\rangle\neq 0$ but $\langle Q\rangle = 0$, we expect protection only for phase errors. When $\langle a^2\rangle = 0$, we expect the memory to become fragile under either noise.}
    \label{fig:phase}
\end{figure}

Finally, the stability of the memory can also be understood as the coexistence of two  order parameters: $\langle Q \rangle = \langle e^{i\pi a^{\dag}a} \rangle \neq 0$ indicates the ferromagnetic phase and therefore suppression of bit-flip errors, while $\langle a^2 \rangle \neq 0$ indicates that the cat states are stabilized, implying suppression of phase-flip errors.
We use a product-state mean-field ansatz $\rho = \bigotimes^M_{x,y=1} \rho_{x,y}$, where each $\rho_{x,y}$ is a density matrix for a two-level system in the basis of $\ket{\pm\alpha_{MF}}$ for some mean-field coherent parameter $\alpha_{MF}$. A non-trivial dissipative phase of the system is identified by non-zero fixed points of $\langle Q\rangle$ and $\langle a^2\rangle$. 
The mean-field solutions suggest that, for small $\kappa_1,\kappa_d$, the memory is protected against both phase and bit-flip errors.
When $\kappa_1$ or $\kappa_d$ exceeds a  threshold, the order parameters undergo two second-order phase transitions and the quantum memory is no longer stable (see the SM \cite{SM}). The mean-field phase diagram is sketched in Fig.~\ref{fig:phase}. 

% \leo{The existence of a dissipative phase suggests that the stability of the memory does not rely on fine-tuning of the parameters.}

\textit{Implementing the photonic-Ising dissipators.}
The key ingredients for our proposal are the microscopic  dissipators defined in Eq.~\eqref{eq:nn}. A direct approach to achieve such terms involves engineering an Ising-like interaction between cavity modes: $H_S \propto - \sum_{\langle i j \rangle} Q_i Q_j$. %At low temperatures, 
%For a sufficiently cold bath, t
The natural system-bath interaction of the form $\sum_i (a_i+a^{\dag}_i)\otimes B_i$ (where $B_i$ acts on bath degrees of freedom) would then give rise to the model described above (within the standard Born-Markov approximation)~\cite{SM}. The  parity-parity interaction $H_S$ can be engineered from coupling between high-impedance cavity modes and Josephson junctions~\cite{cohenthesis,cohen2017}, as we review in the SM~\cite{SM}. %We argue in the SM~\cite{SM} that this physical realization allows us to recover the desired photonic-Ising model with exponentially long lifetime, when each cavity with photon number $N$ is coupled to a chain of $O(N)$ superconducting circuit elements. }

Inspired by the microscopic dissipators Eq.~\eqref{eq:nn}, an alternative approach to protect the memory  involves digitally implementing a stochastic local error decoder. In the SM~\cite{SM}, we provide an explicit description of how to achieve such a local decoder autonomously without the need of measurements; however, it requires  that the local error decoding should be carried out at a rate that scales linearly in the photon number $N$ of each cavity. The implementation can be achieved simultaneously with the dephasing protection Eq.~\eqref{eq:lattice_2ph}, making it fault-tolerant~\cite{prx2019}. Note that this is different from the active repetition cat code in 1D~\cite{prx2019} as we avoid the processing of non-local syndrome information.

\textit{Discussion and outlook.}---We proposed a  photonic-Ising model that hosts robust quantum memory under both single-photon loss and dephasing noise. We can estimate the logical error rates in the photonic-Ising model as follows. While the bit-flip error rate becomes extensive ($\sim O(N)$) in the limit of large cavity photon number, the Ising-type interaction gives rise to an exponentially-suppressed error rate $O(e^{-\gamma M})$ with $\gamma > 0$~\cite{Thomas1989,Schonmann1987,randall:2006}, resulting in a logical bit-flip error rate of $O(Ne^{-\gamma M})$. Similarly, a single cavity yields a phase-flip error rate of $O(e^{-\gamma'N})$ with $\gamma' > 0$, while this is made extensive by the spatially-extended lattice configuration, resulting in a logical phase flip error rate of $O(M^2e^{-\gamma'N})$. Harmonic oscillators with small non-linearities and outstanding coherence properties---and thus with large achievable $N$---can be  found in a variety of photonic and phononic systems (e.g.\ \cite{berdou2022,home11}). 

The realization of the parity-parity coupling based on Josephson junctions and a high-impedance cavity mode is experimentally challenging.  Future efforts should consider other experimental schemes that can lead to the same effective model.
%An open technical question 
%remains to 
%is how to 
%find new routes to achieve 
%methods for the realization which are easier to achieve.
The photonic-Ising model can  be  generalized to adapt the Toom's rule~\cite{toom}, or to  higher dimensions~\cite{poulin:2019}  for a more robust perturbative stability. The full perturbative stability of the model remains an interesting open question.

%The phase-flip error rate is exponentially suppressed in the number of cavity photons. Harmonic oscillators with small non-linearities---and thus with potentially large achievable $N$---can be  found in a variety of photonic and phononic systems. For example, for a single trapped ion with a ground-state wavefunction size of $x_0 \sim 10$ nm and a nonlinearity length scale of $l_\textrm{nl} \sim 1$ mm \cite{home11}, the nonlinearity becomes comparable to the frequency of the oscillator only at $N \sim (l_\textrm{nl}/x_0)^2 \sim 10^{10}$.}

\begin{acknowledgments}

\textit{Acknowledgment.}---We sincerely thank Oles Shtanko, Victor Albert, and Daniel Slichter for useful discussions. We thank Fernando Brand\~{a}o for pointing out Ref.~\cite{poulin:2019} regarding the perturbative stability of a three-dimensional model. S.L.~was supported by the NIST
NRC Research Postdoctoral Associateship. Y.-J.L was supported by the Max Planck Gesellschaft (MPG) through the International Max Planck Research School for Quantum Science and Technology (IMPRS-QST). A.V.G.~acknowledges funding by NSF QLCI (award No.~OMA-2120757), DoE QSA, DoE ASCR Accelerated Research in Quantum Computing program (award No.~DE-SC0020312), ARO MURI, AFOSR, DARPA SAVaNT ADVENT, DoE ASCR Quantum Testbed Pathfinder program (award No.~DE-SC0019040), U.S.~Department of Energy Award No.~DE-SC0019449, NSF PFCQC program, and AFOSR MURI.

\end{acknowledgments}

\bibliography{toric_code.bib}

\begin{thebibliography}{64}
\bibitem [{fn3()}]{fn3}%
  \BibitemOpen
  \href@noop {} {}\bibinfo {note} {We could also include jumps that flip a spin
  if there are two domain walls surrounding it, e.g.~$L^{(2)} = \sqrt{\gamma}
  X_{x,y} P^+_{x,y; \rightarrow} P^+_{x,y; \uparrow} P^-_{x-1,y; \rightarrow}
  P^-_{x,y-1; \uparrow} $, since such a process does not change the energy and
  hence respects detailed balance (for any rate $\gamma$). Consequently, such
  jumps do not change the exact thermal steady state solutions. We choose not
  to include such jumps since the mean-field solution is more accurate without
  them.}\BibitemShut {Stop}%
\bibitem [{fn2()}]{fn2}%
  \BibitemOpen
  \href@noop {} {}\bibinfo {note} {If the symmetry-broken initial state $|
  \bar{0}\rangle$ is recoverable, then so is any state in the code space. This
  is because an arbitrary state in the code space can be expressed as: $c_0 |
  \bar0 \rangle + c_1 | \bar1 \rangle = (c_0-c_1)| \bar0 \rangle + c_1(| \bar0
  \rangle + | \bar1 \rangle) $. Since the dynamics is linear, and the symmetric
  state $| \bar{0}\rangle + | \bar{1}\rangle $ can always be properly recovered
  in the absence of $Z$ dephasing, this implies that any state in the code
  space is recoverable if $| \bar{0}\rangle$ is recoverable.}\BibitemShut
  {Stop}%
\end{thebibliography}
\bibliographystyle{apsrev4-1}

\newpage
\afterpage{\blankpage}

\newpage
\widetext

%%%%%%%%%% Merge with supplemental materials %%%%%%%%%%
%%%%%%%%%% Prefix a "S" to all equations, figures, tables and reset the counter %%%%%%%%%%
\setcounter{equation}{0}
\setcounter{figure}{0}
\setcounter{table}{0}
\setcounter{page}{1}

\renewcommand{\theequation}{S\arabic{equation}}
\renewcommand{\thefigure}{S\arabic{figure}}
%\renewcommand{\bibnumfmt}[1]{[S#1]}
%\renewcommand{\citenumfont}[1]{S#1}
%%%%%%%%%% Prefix a "S" to all equations, figures, tables and reset the counter %%%%%%%%%%

\begin{center}
		{\fontsize{12}{12}\selectfont
			\textbf{Supplemental Material for  ``\papertitle''\\[5mm]}}
		
\end{center}
\normalsize\

The Supplemental Material is organized as follows: In Sec.~1, we provide numerical evidence that the idealized bit flip approximation introduced in the main text is reasonable in the limit of large drive, small single-photon loss, and no dephasing. In Sec.~2, we study a ``toy model'', which was introduced in the main text, which mimics the dynamics of the 2D photonic-Ising model, and which is tractable both numerically and analytically. This model suggests that leakage out of the codespace arising from single-photon loss and dephasing is not detrimental to passive correction. In Sec.~3, we provide details on the mean-field theory order parameters described in the main text. In Sec.~4 we show that the model studied in the main text can be achieved for an Ising-like Hamiltonian interaction between cavity modes in the presence of the natural system-bath coupling. In Sec.~5 we describe a way to achieve an Ising-like interaction between cavity modes in a superconducting circuit scheme. In Sec.~6, we provide an alternative way of achieving the model in the main text; instead of engineering a Hamiltonian interaction, we describe a way to engineer only the desired dissipators. This scheme relies on the presence of fast unitary gates and ancilla resets. 

\section{1.~Idealized bit flip approximation}\label{sm:sec:idealized_flip}

In this section, we elaborate on the idealized bit flip approximation used in the main text. In  experiments, the bit flip error  for a single photonic cat qubit is generated via single-photon loss $L_1 = \sqrt{\kappa_1} a$. However, in order to map our many-body-cat-qubit system to the 2D Ising model, we must replace this noise generator with an ``idealized bit flip'', represented via the jump operator: $E_1 = \sqrt{\kappa_1} a V$ where $V$ is a projector onto the codespace. We provide evidence that $E_1$ is a reasonable approximation for $L_1$ in the limit of  small single-photon loss and large two-photon drive (compared to the two-photon loss rate), which is the  relevant regime for  modern experiments involving  photonic cat qubits \cite{amazon_cat}. We also assume the absence of photon dephasing.  To this end, we shall present two models for a single cavity and show that their steady states and dissipative gaps converge in this limit. 
  
 Model 1 has the   standard single-photon loss term which is expected  to appear in experiment.  Model 2 has the ``idealized bit flip'' which is needed to make numerical progress.

\textit{Model 1}: Let us consider a single photonic cavity in the presence of two-photon drive $H = \lambda [a^2 + (a^\dagger)^2]$, two photon loss $L_2  = \sqrt{\kappa_2} a^2$, and single-photon loss $L_1 = \sqrt{\kappa_1} a$. It is convenient to utilize the gauge freedom of the Lindbladian to eliminate the Hamiltonian by incorporating it in a dissipative term. The following two dissipators share the same master equation as the model  just described:
\begin{align} \label{eq:bit_flip_dis1}
L_c &= \sqrt{\kappa_2} (a^2 - \alpha^2), \qquad \alpha = \sqrt{\frac{\lambda}{\kappa_2} } e^{-i \pi/4} \\
 L_1  &= \sqrt{\kappa_1} a.  \label{eq:bit_flip_dis2}
\end{align}
The dissipator $L_c$ will cause states in the Hilbert space to evolve towards the coherent states $| \pm \alpha \rangle$, which are dark states of $L_c$.   We thus find that $L_c$ generates the ``recovery'' part of the Lindbladian, while $L_1$ generates bit flip errors and causes leakage out of the codespace. 

From the perspective of quantum trajectories, single-photon loss causes the amplitude of a coherent state to decay due to the non-Hermitian Hamiltonian term  proportional to  $ \kappa_1 a^\dagger a$ which (by itself) causes the coherent state parameter to decay via $\alpha e^{-\kappa_1 t}  $. The two-photon drive process ensures that the steady state amplitude remains non-zero, but nevertheless the photon population decreases due to the single-photon loss.  Within mean-field theory, the  average number $\bar{n}$ of photons in the cavity  satisfies
\begin{equation}
\bar{n} = \frac{2 \lambda - \kappa_1}{2\kappa_2}. 
\end{equation}
 This suggests  that, in the limit of $\lambda / \kappa_2 \gg 1$,  the steady state of the system should start to converge to a coherent state $ | \pm \mu \rangle$ with a shifted amplitude:
\begin{equation}
a | \pm\mu \rangle =  \pm \mu | \pm\mu \rangle, \qquad  \mu=  \sqrt{ \frac{2 \lambda - \kappa_1}{2\kappa_2} } e^{-i \pi/4}.
\end{equation}
Numerics suggest that the true steady state of the system will be a  mixture of several pure states \cite{lieu2020}. However,  the steady state will have large overlap with the states  $ |  \pm\mu \rangle $.  In the limit $\kappa_1 / \kappa_2 \ll 1$, the steady state will start to converge to  a mixture of the  states $ | \pm\mu \rangle$.

\begin{figure}
    \centering
          \includegraphics[scale=0.3]{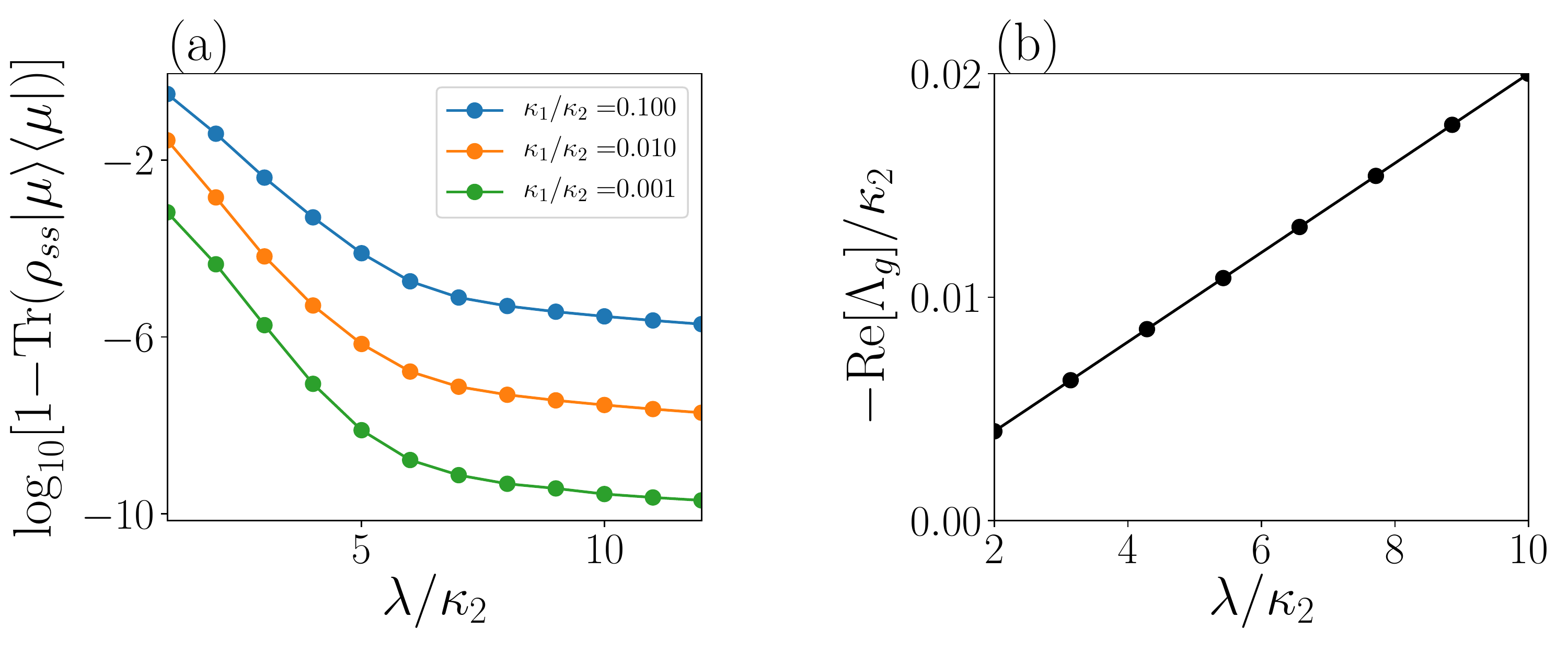} 
    \caption{Model 1: (a) Expectation value of $|\mu \rangle \langle \mu |$ in the steady state of the model described in Eqs.~\eqref{eq:bit_flip_dis1}, \eqref{eq:bit_flip_dis2}  with  $\lambda / \kappa_2 = N$ for different choices of $\kappa_1 / \kappa_2$. In the limit $\lambda / \kappa_2 \gg 1, \kappa_1 / \kappa_2 \ll 1$, the system converges to the coherent state $|\mu \rangle$. We use exact Lindblad evolution starting from the initial state $|\alpha \rangle$ and evolving for a time $t = 200/ \kappa_2$ to reach the steady state. (b)  The dissipative gap $\Lambda_g$ scales linearly as a function of the drive strength, for $\kappa_1 / \kappa_2 = 10^{-3}$.}
    \label{fig:bitflip1}
\end{figure}

We can confirm this via numerical simulations. In Fig.~\ref{fig:bitflip1} we plot the overlap of the steady state with $| \mu \rangle$ as a function of the drive strength $\lambda/ \kappa_2$, for different choices of $\kappa_1 / \kappa_2$. We find that the steady state of the system  approaches $| \mu \rangle$ in the limit $\lambda / \kappa_2 \gg 1, \kappa_1 / \kappa_2 \ll 1$. These parameters are in a regime that is relevant  for modern experiments \cite{amazon_cat}. We also plot the dissipative gap, which scales linearly with the drive strength. 

\begin{figure}
    \centering
          \includegraphics[scale=0.3]{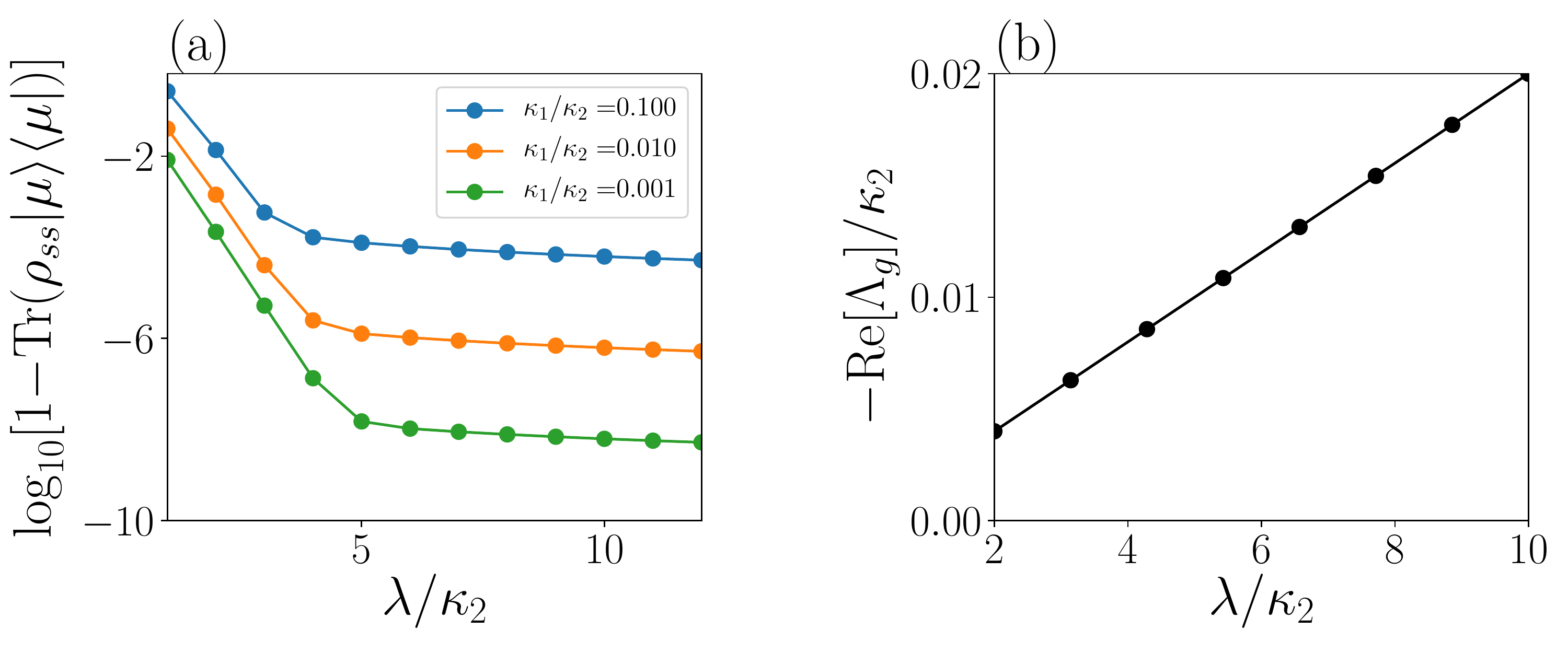}

    \caption{Model 2: (a) Expectation value of $|\mu \rangle \langle \mu |$ in the steady state of the model described in Eqs.~\eqref{eq:bit_flip_dis3}, \eqref{eq:bit_flip_dis4}  with parameters: $\lambda / \kappa_2 = N$ for different choices of $\kappa_1 / \kappa_2$. In the limit, $\lambda / \kappa_2  \gg 1, \kappa_1 / \kappa_2 \ll 1$ the system converges to the coherent state $|\mu \rangle$. We use exact Lindblad evolution starting from the initial state $|\alpha \rangle$ and evolving for a time $t = 200/ \kappa_2$ to reach the steady state. (b)  The dissipative gap $\Lambda_g$ scales linearly as a function of the drive strength, for $\kappa_1 / \kappa_2 = 10^{-3}$.}
    \label{fig:bitflip2}
\end{figure}

Beyond a shift in the coherent state amplitude, single-photon loss also has the effect of reducing the qubit-steady-state structure to a classical-bit-steady-state structure. Only classical mixtures of coherent states are stable, while off-diagonal coherences have a finite lifetime:
\begin{equation}
\rho_{ss} \approx c |\mu \rangle \langle \mu | + (1-c) |- \mu \rangle \langle - \mu |.
\end{equation}
for $c \in [0,1], \lambda/\kappa_2 \gg 1, \kappa_1 / \kappa_2 \ll 1$. The steady state  is thus two dimensional, enough only to store a classical bit.

\textit{Model 2}:  Let us now consider  a different model which will have  the same steady state  and dissipative gap  in the limit $\lambda/\kappa_2 \gg 1, \kappa_1 / \kappa_2 \ll 1$, but will involve the ``idealized bit flip'' rather than single-photon loss. Consider the dissipators 
\begin{align} \label{eq:bit_flip_dis3}
L_c &= \sqrt{\kappa_2} (a^2 - \alpha^2), \qquad  \alpha= \sqrt{ \frac{ \lambda }{\kappa_2} } e^{-i \pi/4} \\
 E_1  &= \sqrt{\kappa_1} b = \sqrt{\kappa_1} a V, \qquad V = | \alpha_e  \rangle \langle \alpha_e | + | \alpha_o  \rangle \langle \alpha_o | \label{eq:bit_flip_dis4}
\end{align}
where $|\alpha_e  \rangle  \sim |\alpha  \rangle +| -\alpha  \rangle,  | \alpha_o  \rangle  \sim | \alpha \rangle -| -\alpha  \rangle $. In this model, the dissipator $E_1$ does not cause any leakage of photons out of $|\alpha \rangle$. This is because the non-Hermitian Hamiltonian term proportional to $E_1^\dagger E_1$ keeps superpositions of $|\pm \alpha \rangle$ in this subspace (due to the projector $V$). Nevertheless, the term $E_1$ ensures that quantum superpostions of $|\pm \alpha \rangle$ are unstable, while classical mixtures are stable.  The steady state starts to converge to the following state in the limit of large drive $\lambda / \kappa_2 \gg 1$:
\begin{equation}
\rho_{ss} \approx c |\alpha \rangle \langle \alpha | + (1-c) |- \alpha \rangle \langle - \alpha |,
\end{equation}
for $c \in [0,1]$. 

The overlap between $|\alpha \rangle$ and $|\mu \rangle$ satisfies 
\begin{equation}
    |\langle \alpha | \mu \rangle |^2 = \exp[-\frac{\kappa_1^2}{16 \kappa_2 \lambda} ] \approx 1   -\frac{\kappa_1^2}{16 \kappa_2 \lambda} + \ldots 
\end{equation}
 This implies that the deviation from unity scales  as $\kappa_1^2$ when $\kappa_2\lambda \gg \kappa_1^2$.  We confirm this in Fig.~\ref{fig:bitflip2}: The deviation between the steady state of Model 2 and $|\mu \rangle$ scales quadratically with $\kappa_1$ in the limit of large drive. We also plot the dissipative gap, which again scales linearly with the drive strength.

We have shown that  Models 1 and 2 converge to each other in terms of their steady state and their dissipative gap in the limit $\lambda / \kappa_2 \gg 1, \kappa_1 / \kappa_2 \ll 1$.  This suggests that Model 2 is a reasonable approximation for Model 1 in this regime. Intuitively, this happens because the system quickly evolves toward the codespace, such that the projector term $V$ acts trivially on the state. In the main text, we demonstrated that Model 2 passively corrects against bit flip errors via the Ising-like dissipators described above. We expect Model 1 to behave in qualitatively the same manner after the replacement of $b \rightarrow a$.

We  note that, although we used the limit $\lambda / \kappa_2 \gg 1, \kappa_1 / \kappa_2 \ll 1$ to establish the exact mapping to the Ising model, we do not expect that this limit is needed to preserve quantum information in general. Rather,  the system only needs to   stay within the ordered phase (see Fig.~3 in the main text and SM Sec.~3). A relatively small $\kappa_1$ ensures that the steady state of the dynamics is a mixed state. Nevertheless, we expect that this mixed state will be a  ``noiseless subsystem'', which implies that it can be decoded with a channel superoperator at the end of the dynamics.

\section{2.~Toy model}\label{sm:sec:toy}

The Ising-inspired bit-flip recovery jump operators  [Eqs.~\eqref{eq:nn} in the main text] by themselves will not give rise to protection against single-photon loss in the absence of a drive, since single-photon loss will cause the system to evolve to a vacuum state. In this section, we argue that, when the bit-flip recovery is coupled with the driving, the resulting environment is able to protect against both dephasing and single-photon loss errors.

Ideally, we would like to numerically simulate the 2D array of $M^2$ cat qubits introduced in the main text. However,  such a simulation is computationally expensive. We restrict ourselves to the toy model introduced in the main text: a single cat qubit coupled to a two-level system, the latter described by Pauli operators $X, Y, Z$. The logical states of this toy system are defined as $\ket{\downarrow}\ket{\alpha_o}$ and $\ket{\downarrow}\ket{\alpha_e}$, where $\ket{\alpha_e},\ket{\alpha_o}$ are the logical states for a single cat qubit. The noise and recovery jump operators are modified to
\begin{align}
l_c &= \sqrt{\kappa_2} (a^2 - \alpha^2), \qquad \alpha = \sqrt{\frac{\lambda}{\kappa_2} } e^{-i \pi/4} \\
 l_1  &= \sqrt{\kappa_1} X a,\quad l_d = \sqrt{\kappa_d} a^{\dag}a,  \\
 l_{nn} &= \sqrt{\kappa_{nn}} \frac{1}{2}X(1-Z) a,
\end{align}
where $l_c$ generates a Lindbladian that is equivalent to the combined action of $h$ and $l_2$ in the main text.
In this toy model, the spin-$1/2$ particle is essentially a ``classical bit'' that takes the discrete value of up or down. Any single-photon loss event is always accompanied by a flip of the spin. A bit-flip recovery for the cat qubit can then be achieved by checking the orientation of the spin: an annhilation operator $a$ is applied to the cavity if the spin points upwards, otherwise nothing happens. This mimics the full 2D case where a bit-flip recovery jump is triggered by a parity misalignment between nearest-neighbor cat qubits. The difference between the 2D model and the toy model is that the latter always knows when an odd number of single photon-loss events has occurred. What remains to be tested is whether the errors can be corrected by introducing the bit-flip recovery jump. 

\begin{figure}
    \centering
    \includegraphics[scale = 0.33]{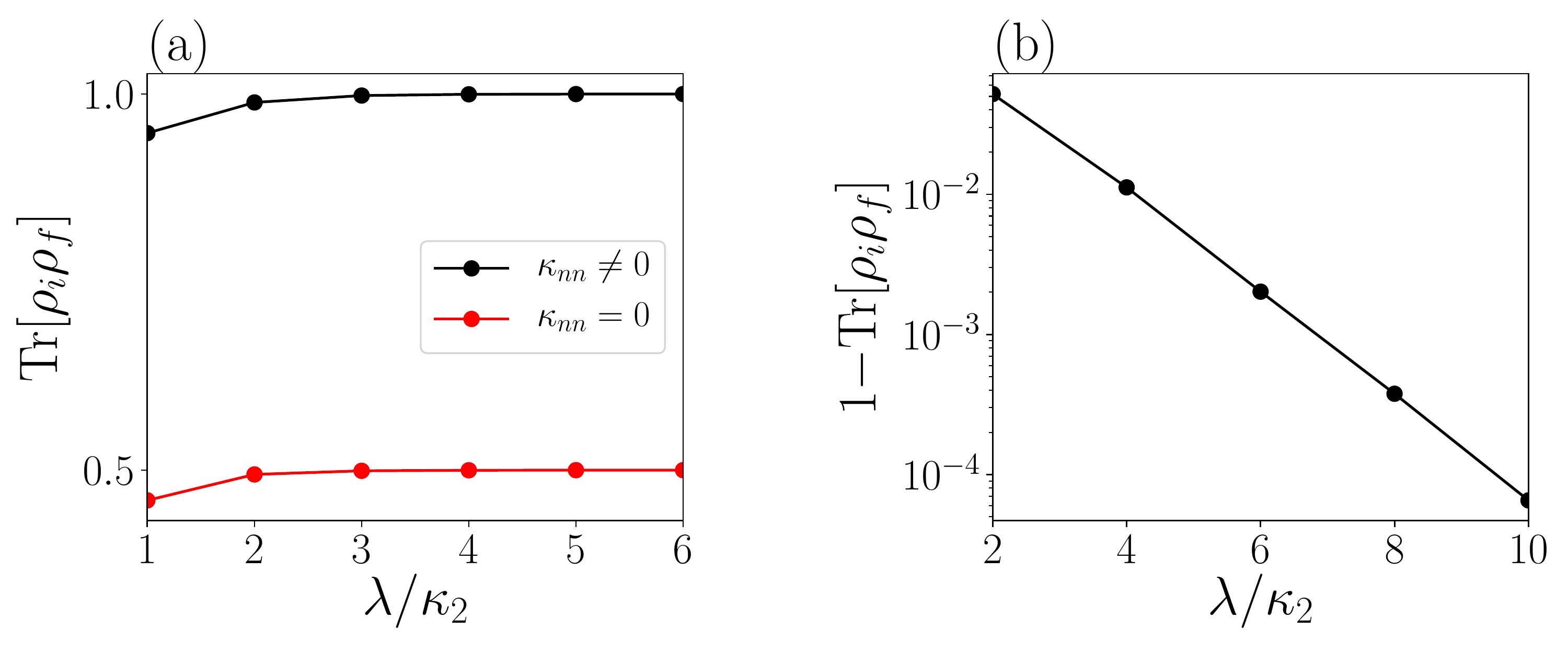}
    \caption{We initialize the dynamics with state $\rho_i = |\psi \rangle \langle \psi |$, where $|\psi \rangle = \frac{1}{\sqrt{5}}\ket{\downarrow}\ket{\alpha_e}+\frac{2e^{i\pi/4}}{\sqrt{5}}\ket{\downarrow}\ket{\alpha_o}$. (a) Overlap between the initial and final states for $\kappa_{nn} = 0$ [case (i)] and $\kappa_{nn}/\kappa_2 = 0.3$ [case (ii)] as $N = \lambda/\kappa_2$ increases. Parameters: $\kappa_d/\kappa_2 = 0.1,\kappa_{1}/\kappa_2 = 0.1$. (b) For case (ii), i.e.\ $\kappa_{nn}\neq 0$, the log scale plot shows that the fidelity converges exponentially quickly to 1 as $N\to\infty$.}
    \label{sm:fig:bitflip}
\end{figure}

Suppose we initialize the dynamics with a generic state in the codespace. 
We consider the following two scenarios: We choose the model with (i) $\kappa_2 = 1, \kappa_d = 0.1,\kappa_{1} = 0.1,\kappa_{nn} = 0$ and (ii) $\kappa_2 = 1, \kappa_d = 0.1,\kappa_{1} = 0.1,\kappa_{nn} = 0.3$. The system size parameter is $N = \lambda/\kappa_2$ with $N\to\infty$ representing the thermodynamic limit. The initial state is first evolved with this Lindbladian for duration $T = 15$, then followed by the corresponding noiseless Lindbladian evolution ($\kappa_d = \kappa_1 = 0$) for another $T = 15$. In the end, we compute the fidelity between the final state and the initial state. The results for the two scenarios are shown in Fig.~\ref{sm:fig:bitflip} for different $N$.

The results clearly show distinct behaviors. For case (i), where $\kappa_{nn} = 0$, the single-photon loss causes uncorrectable errors in the stored memory, leading to a saturated fidelity of $1/2$ (due to an equal mixture of the flipped and unflipped states) as $N$ increases. For case (ii), where $\kappa_{nn}\neq 0$, increasing $N$ leads to a fidelity exponentially close to the ideal value of 1.

\begin{figure}
    \centering
    \includegraphics[scale = 0.33]{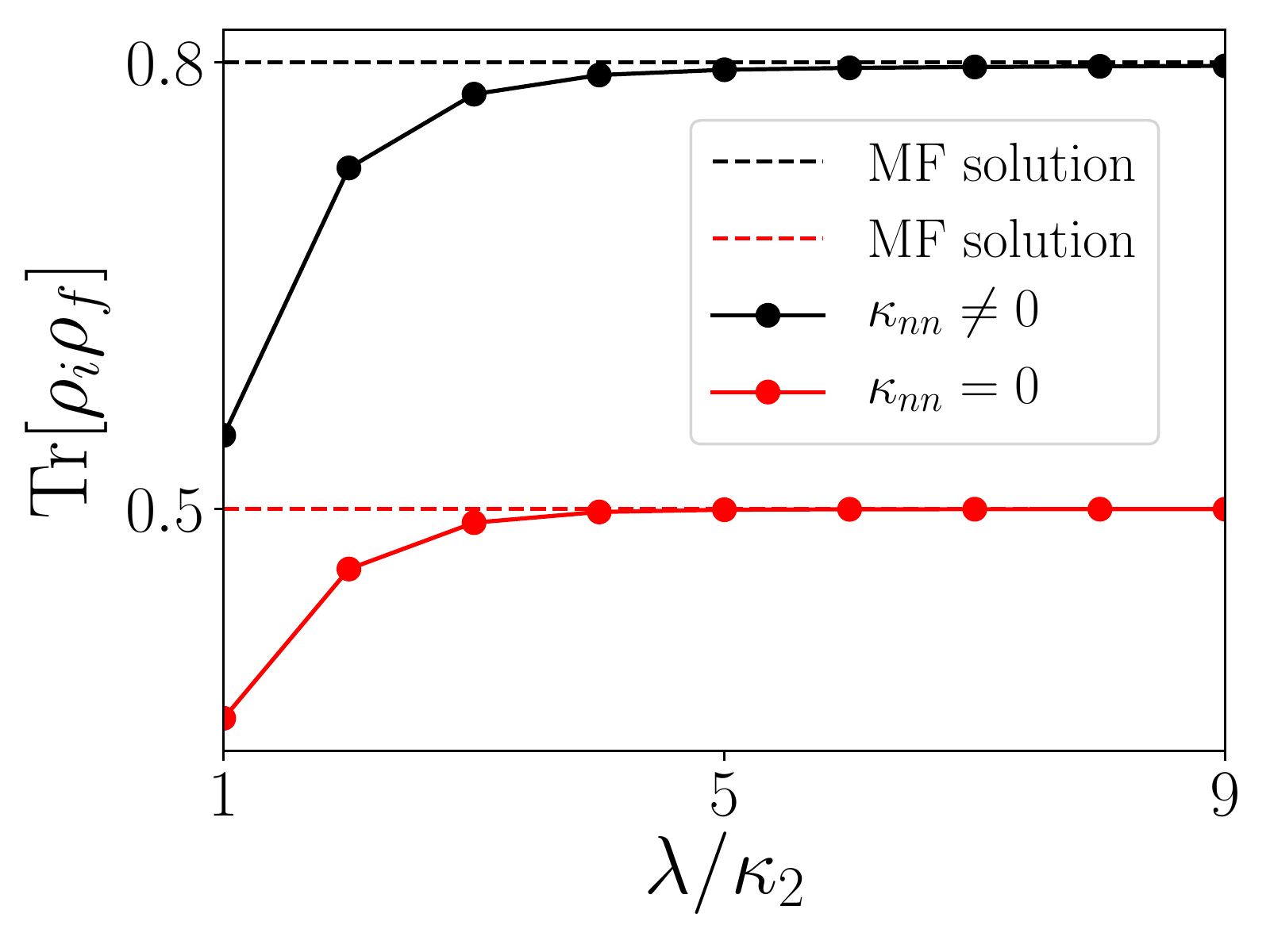}
    \caption{ Repeating the same simulation as in Fig.~\ref{sm:fig:bitflip}, except the recovery (noiseless) Lindblad evolution is done using $\kappa_{nn} = \kappa_d = \kappa_1 = 0$ in both cases (i) and (ii) [not just case (i)].  The overlap between the initial and final states is larger than $1/2$ when $\kappa_{nn}\neq 0$ during the noisy dynamics [modified case (ii)], while the overlap saturates to  $1/2$  when $\kappa_{nn} = 0$ [the original case (i)]. In the thermodynamic limit, the overlap values agree with   mean-field results (shown as horizontal dashed lines).}
    \label{sm:fig:bitflip2}
\end{figure}

As a sanity check, let us consider the same numerical simulation but modify case (ii) by setting $\kappa_{nn}$ = 0 during the noiseless dynamics (while still keeping $\kappa_{nn} = 0.3$ during the noisy dynamics). 
The results of the simulation are shown in Fig.~\ref{sm:fig:bitflip2}. In case (i), which is identical to the one studied in Fig.~\ref{sm:fig:bitflip}, the fidelity relaxes to $1/2$ regardless of the system size as before. The modified case (ii) shows a saturated fidelity between $1/2$ and $1$, suggesting a partial preservation of the initial quantum memory. This again confirms the dynamical quantum memory protection arising from the flip-recovery jump and two-photon drive. 


\textbf{Mean-field analysis of the toy model.}---We use a mean field approach to show that, despite the spin-boson coupling in our toy model, the $ \mathbb{Z}_2$ symmetry-breaking phase diagram of the single cat qubit is reproduced. Given an observable $\hat O$ and a Lindbladian term $\hL$ generated by the jump operator $L$, the expectation value obeys
\begin{equation}\label{sm:eq:expval}
    \Tr[\hat O\hL\rho] = -\frac{1}{2}\Tr\left[[\hat O,L^{\dag}]L\rho+L^{\dag}[L,\hat O]\rho\right].
\end{equation}
Using this, we can derive a coupled set of mean-field equations of motion for $\langle  a\rangle$ and $\langle  Z\rangle$:
\begin{align}
    \frac{d}{dt}\langle  a\rangle&= 
    -i\lambda\langle a^{\dag}\rangle - \frac{1}{2}\left(\kappa_1+\kappa_d+\frac{\kappa_{nn}}{2}(1-\langle Z\rangle)\right)\langle  a\rangle-\kappa_2|\alpha|^2\langle  a\rangle, \\
     \frac{d}{dt}\langle  Z\rangle&= -2\kappa_1 |\alpha|^2\langle Z\rangle + \kappa_{nn}|\alpha|^2(1-\langle Z\rangle).
\end{align}
This yields the mean-field fixed point solutions for both observables
\begin{align}
    \langle Z\rangle &= \frac{\kappa_{nn}}{\kappa_{nn}+2\kappa_1}, \\
    \kappa_2|\alpha|^2 &= 
    |\lambda|-\frac{1}{2}\left(\kappa_1+\kappa_d+\frac{\kappa_1\kappa_{nn}}{\kappa_{nn}+2\kappa_1}\right).
\end{align}
The expression  closely matches   the simulation in the thermodynamic limit (see Fig.~\ref{sm:fig:bitflip2}). 

It is interesting to note that if $\kappa_1 / \kappa_2$ is small enough, then any non-zero $\kappa_{nn}$ can give rise to a stable memory ($\langle Z\rangle,\langle a\rangle\neq 0$). On the other hand, if $\kappa_1 / \kappa_2$ is large, a large $\kappa_{nn}$ can destabilize the memory, leading to $\langle a\rangle = 0$.

\section{3.~Mean-field solution for the 2D photonic-Ising model}\label{sm:sec:Ising_cat_mf}
In this section, we present the mean-field solution for the 2D photonic-Ising model. The mean-field analysis shows the existence of two symmetry-breaking transitions via two order parameters: $a^2$ and $Q\equiv e^{i\pi  a^{\dag}a}$. 

We consider a product-state mean-field ansatz $\rho = \bigotimes^M_{x,y=1} \rho_{x,y}$. At each site, $\rho_{x,y}$ is a density matrix for a two-level system in the basis of $\ket{\pm \alpha_{MF}}$ for some coherent parameter $\alpha_{MF}$. 
We first begin by deriving the mean-field equation for $Q = e^{i\pi a^{\dag}a}$. Note that all the terms  that commute with $Q$
do not contribute to the time evolution. We are therefore left to consider only the single-photon loss term and the bit-flip correction term. Using Eq.~\eqref{sm:eq:expval}, we obtain
\begin{align}
    \frac{d\langle Q\rangle}{dt} = -2\left(\kappa_1\langle a^{\dag}a Q\rangle +\kappa_{nn}\langle a^{\dag}a QP_{\kappa_{nn}}\rangle+\tilde{\kappa}_{nn}\langle a^{\dag}a QP_{\tilde{\kappa}_{nn}}\rangle\right),
\end{align}
where $P_{\kappa_{nn}},P_{\tilde{\kappa}_{nn}}$ are sums of projectors onto different parity configurations with rates $\kappa_{nn},\tilde{\kappa}_{nn}$, as introduced in the main text. Within  mean-field theory, we replace the expectations by a product of expectations at each site, yielding
\begin{align}
    -\frac{1}{2|\alpha|^2}\frac{d\langle Q\rangle}{dt}=
    \frac{\kappa_{nn}-4\tilde{\kappa}_{nn}}{16}\langle Q\rangle^5+\frac{\kappa_{nn}+4\tilde{\kappa}_{nn}}{8}\langle Q\rangle^3-\left(\frac{3\kappa_{nn}+4\tilde{\kappa}_{nn}}{16}-\kappa_1\right)\langle Q\rangle .
\end{align}
Similarly, we can derive the mean-field equation for $a^2$:
\begin{equation}
    \frac{d\langle a^2\rangle}{dt}=
    -\kappa_2(2\langle a^{\dag}aa^2\rangle +\langle a^2\rangle) -i\lambda(2\langle a^{\dag}a\rangle +1)-\kappa_1\langle a^2\rangle -2\kappa_d\langle a^2\rangle-\kappa_{nn}\langle a^2 P_{\kappa_{nn}}\rangle -\tilde{\kappa}_{nn}\langle a^2 P_{\tilde{\kappa}_{nn}}\rangle.
\end{equation}
With the mean-field ansatz, we may approximate $\langle a^{\dag}aa^2\rangle \approx |\alpha_{MF}|^2\langle a^2\rangle$. We also have $\langle a^2 P_{\tilde{\kappa}_{nn}}\rangle = \langle a^2 \rangle\langle P_{\tilde{\kappa}_{nn}}\rangle$ and $\langle a^2 P_{\kappa_{nn}}\rangle = \langle a^2 \rangle\langle P_{\kappa_{nn}}\rangle$.
After some algebra, the mean-field fixed points at the thermodynamic limit (e.g. $\kappa_2\to 0$) can be found to satisfy
\begin{align}
    \langle Q\rangle^2 &=  \frac{2\sqrt{\kappa_{nn}^2-4\kappa_1(\kappa_{nn}-4\tilde{\kappa}_{nn})}- \kappa_{nn}-4\tilde{\kappa}_{nn}}{\kappa_{nn}-4\tilde{\kappa}_{nn}},
    \\
   |\alpha_{MF}|^2 &=
    \frac{2\lambda -\kappa_1-2\kappa_d-\gamma_4 \langle Q\rangle^4-\gamma_2 \langle Q\rangle^2-\gamma_0}{2\kappa_2}, \label{eq:alpha}
\end{align}
where $\gamma_4 = (-3\kappa_{nn}+4\tilde{\kappa}_{nn})/16$,  $\gamma_2 = (\kappa_{nn}-4\tilde{\kappa}_{nn})/8$, and $\gamma_0 = (\kappa_{nn}+4\tilde{\kappa}_{nn})/16$. In addition, $\langle Q\rangle^2\neq 0$ is only possible when $|\alpha_{MF}|^2 \neq 0$. Intuitively, when $\langle a^2\rangle = 0$, the cavity will lose coherence and decay to the vacuum due to the noise. The logical states are no longer well-defined in this case.

It is important to note that the mean-field solution suggests that the leakage caused by both finite $\kappa_1$ and finite $\kappa_d$ is compensated by the two-photon drive. %Their effects 
The effect of this leakage amounts to a shift in the steady state coherent parameter.

\section{4.~A microscopic derivation of the photonic-Ising dissipators}

Here we establish an explicit connection between the Hamiltonian approach for the photonic-Ising model and the microscopic Lindbladian approach. This section (Section 4) provides an example where  the proposed photonic-Ising dissipators emerge naturally from a microscopic coupling, unlike the example in the previous section. 
In the next section (Section 5), we will discuss an experimental protocol that realizes the desired Hamiltonian coupling based on superconducting circuits.

\subsection{The microscopic generators}
Let us start by considering a microscopic Hamiltonian of both the system and the bath:
\begin{equation}
    H = H_S + H_B + H_{SB},
\end{equation}
where $H_S = -J\sum_{\langle i,j\rangle}Q_iQ_j$, and $H_B,H_{SB}$ are the bath and the system-bath coupling Hamiltonian, respectively. Notice that, in contrast to the main text, we explicitly introduced an energy scale $J>0$ in $H_S$ to help us carry out the analysis. We consider $H_{SB}=\sum_i (a_i+a^{\dag}_i)\otimes B_i $, where $a_i$ is the annihilation operator on the photonic-Ising lattice and $B_i$ is some Hermitian local operator on the bath.  We assume that the bath is large and the interaction $H_{SB}$ is weak such that effects of the coupling on the bath is fast and can be neglected, i.e.\ the full density matrix approximately factorizes into a product of a system density matrix and a bath density matrix: $\rho(t) \approx \rho_S(t)\otimes \rho_B,\ \forall t$. Provided that the standard Born-Markov approximation (i.e.\ the smallness of the influence of the system-bath coupling on the bath) and the rotating-wave approximation are valid~\cite{breuer2002}, we can derive the Master equation in the interaction picture as
\begin{equation}\label{sm:eq:master_eq}
    \frac{d\rho_S}{dt}= -i[H', \rho_S]+ \sum_{\omega}\sum_{i,j}\gamma_{i,j}(\omega)\left( A_{i}(\omega)\rho_S A_{j}^{\dag}(\omega)-\frac{1}{2}\{A^{\dag}_{j}(\omega)A_{i}(\omega),\rho_S\}\right).
\end{equation}
Here $H'$ is the Lamb-shift Hamiltonian which we will define below. The operator $A_i(\omega)$ is defined as
\begin{equation}
    A_i(\omega) = \sum_{E}\Pi(E)(a_i+a^{\dag}_i)\Pi(E+\omega),
\end{equation}
where $\Pi(E)$ is a projection the eigenstates of $H_S$ of energy E. So $A_i(\omega)$ is a lowering operator: it is the part of $a_i+a^{\dag}_i$ that couples eigenstates of $H_S$ whose energies differ by $\omega$. It is easy to verify that $\sum_{\omega}A_i(\omega) = a_i+a^{\dag}_i$. Note that $a_i+a^{\dag}_i$ can only create an energy difference of $\omega = 0,\pm 4J,\pm 8J$, and we can work out $A_i(\omega)$ for each case explicitly. 
It is straightforward to verify that $A_i(\omega)$ is geometrically local and takes the form
\begin{align}
A_i(0) &= (a_i+a^{\dag}_i)\sum_{\sigma(2)} P_{\sigma(2)},\ 
   A_i(0) = (a_i+a^{\dag}_i)\sum_{\sigma(2)} P_{\sigma(2)},\label{sm:eq:A_0}\\
    A_i(+4J) &= (a_i+a^{\dag}_i)\sum_{\sigma(3)} P_{\sigma(3)},\ 
    A_i(-4J) = (a_i+a^{\dag}_i)\sum_{\sigma(1)} P_{\sigma(1)},\label{sm:eq:A_4}\\
    A_i(+8J) &= (a_i+a^{\dag}_i)\sum_{\sigma(4)} P_{\sigma(4)},\ 
    A_i(-8J) = (a_i+a^{\dag}_i)\sum_{\sigma(0)} P_{\sigma(0)},\label{sm:eq:A_8}
\end{align}
where $P_{\sigma(n)}$ denotes the projector onto different local configurations $\sigma(n)$ around site $i$ with $n$ domain walls. The following relationship is satisfied: $A_i^{\dag}(\omega) = A_i(-\omega)$.

In Eq.~\eqref{sm:eq:master_eq}, the Hamiltonian $H'$ is the Lamb-shift Hamiltonian
\begin{equation}
    H' = \sum_{\omega}\sum_{i,j}\chi_{i,j}(\omega)A^{\dag}_i(\omega)A_j(\omega),
\end{equation}
where the coupling $\chi_{i,j}(\omega)$ is given in terms of the Fourier transform of the reservoir correlation functions~\cite{breuer2002}:
\begin{equation}
    \chi_{i,j}(\omega) = \Im{C_{i,j}(\omega)},\ \text{where }C_{i,j}(\omega)\coloneqq \int^{\infty}_0 ds e^{i\omega s}\langle B^{\dag}_i(s) B_j(0)\rangle,
\end{equation}
where $B^{\dag}_i(s)$ denotes Heisenberg evolution under $H_B$ for time $s$ and the expectation value is taken in the initial thermal state of the bath. To simplify the expression, we make the assumption that $\langle B^{\dag}_i(s) B_j(0)\rangle\approx \langle B^{\dag}_i(s)\rangle \langle B_j(0)\rangle$ for $i\neq j$. In particular, this is immediately satisfied if each site is coupled to its own bath. Note that we also assumed $\langle B^{\dag}_i(s)\rangle =\langle B_j(0)\rangle =0$ in order to get to Eq.~\eqref{sm:eq:master_eq} (see Ref.~\cite{breuer2002}). We find $\chi_{i,j}(\omega) = 0$ for $i\neq j$. Defining $\chi_{i,i}(\omega)\coloneqq \chi_i(\omega)$, $H'$ simplifies to
\begin{align}
    H' &= \sum_{\omega,i}\chi_{i}(\omega)A^{\dag}_i(\omega)A_i(\omega) \nonumber \\
    &= \sum_i\left(\sum_{\omega}\chi_i(\omega)\sum_{\sigma(n[\omega])}P_{\sigma(n[\omega])}\right)(a_i+a_i^{\dag})^2,
\end{align}
where $n[\omega]$ is the number of domain walls in the projected configuration as in Eqs.~\eqref{sm:eq:A_0}-\eqref{sm:eq:A_8}.
We therefore have a Hamiltonian contribution quadratic in the bosonic operators. A Hamiltonian of this form will in general cause dephasing of the quantum memory. However, such an effect will be exponentially suppressed by the combination of two-photon drive and two-photon loss in Eq.\ (\ref{eq:lattice_2ph}) of the main text.

In Eq.~\eqref{sm:eq:master_eq}, the dissipative rate $\gamma_{i,j}(\omega)$ is also expressed in terms of the correlation functions as $\gamma_{i,j}(\omega) = \Re[C_{i,j}(\omega)]=\int^{\infty}_{-\infty} ds e^{i\omega s}\langle B^{\dag}_i(s) B_j(0)\rangle$. We make use of the locality assumption again so $\gamma_{i,j} =0$ for $i\neq j$.
The master equation Eq.~\eqref{sm:eq:master_eq} simplifies to
\begin{align}\label{sm:eq:master_eq_simple}
 \frac{d\rho_S}{dt}= -i[H', \rho_S]+
    \sum_{\omega\in\{0,\pm 2J,\pm 4J\}}\sum_{i}\gamma_{i}(\omega)\left( A_{i}(\omega)\rho_S A_{i}^{\dag}(\omega)-\frac{1}{2}\{A^{\dag}_{i}(\omega)A_{i}(\omega),\rho_S\}\right). 
\end{align}
Imposing the  Kubo-Martin-Schwinger (KMS) condition (i.e.~assuming that the bath is in thermal equilibrium \cite{breuer2002}) leads to the detailed balance relation
\begin{equation}
    \gamma_i(\omega) = e^{\beta\omega}\gamma_i(-\omega).
\end{equation}
for some temperature $\beta = 1/(k_BT)$ set by the bath. 

\subsection{Recovering the photonic-Ising dissipators}
We now show that 
the photonic-Ising dissipators defined in Eq.\ (\ref{eq:nn}) of the main text are a special case of the dissipative part of Eq.~\eqref{sm:eq:master_eq_simple}. 
Let us set $\gamma(0) = \gamma(-4J) = \gamma(-8J) = \kappa_1/2$. This can be satisfied at low temperature for a roughly constant density of states across this energy range. 
Suppose we initialize the system $\rho_S(0)$ as a pure state corresponding to a well-defined domain-wall configuration. Let us denote $\mathcal{D}(L)[\rho_S] = L\rho_SL^{\dag}-\frac{1}{2}\{L^{\dag}L,\rho_S\}$, then, for any site $i$ and time $t$, the density matrix $\rho_S(t)$ satisfies
\begin{equation}
    \mathcal{D}\left(a_i\sum_{\sigma}P_{\sigma}\right)[\rho_S] = \sum_{\sigma}\mathcal{D}(a_iP_{\sigma})[\rho_S],
\end{equation}
where the sum is taken over some projector $P_{\sigma}$ that projects onto distinct domain wall configuration $\sigma$. The same result holds when we replace $a_i$ by $a_i^{\dag}$ or $a_i+a_i^{\dag}$. This is because $\rho_S(t)$ evolved under Eq.~\eqref{sm:eq:master_eq_simple} is always a linear combination of some $\ketbra{\psi_1}{\psi_2}$, where $\ket{\psi_1},\ket{\psi_2}$ might be different states but they are always the same domain wall configuration defined by the parity misalignment. The dissipative terms on site $i$ in Eq.~\eqref{sm:eq:master_eq_simple} now read
\begin{align}
    &\gamma_i(0)\mathcal{D}(A_i(0))[\rho_S]
    + \sum_{\omega = \pm 4J}\gamma_i(\omega)\mathcal{D}(A_i(\omega))[\rho_S]
    +  \sum_{\omega = \pm 8J}\gamma_i(\omega)\mathcal{D}(A_i(\omega))[\rho_S] \nonumber \\
    &= \frac{\kappa_1}{2} \sum_{\omega = 0,\pm 4J, \pm 8J}\mathcal{D}(A_i(\omega))[\rho_S] + \left(\gamma_i(2J)- \frac{\kappa_1}{2}\right)\mathcal{D}(A_i(4J))[\rho_S]+
    \left(\gamma_i(8J)- \frac{\kappa_1}{2} \right)\mathcal{D}(A_i(8J))[\rho_S] \nonumber \\
     &= \frac{\kappa_1}{2}  \mathcal{D}\left( \sum_{\omega = 0,\pm 4J, \pm 8J}A_i(\omega)\right)[\rho_S]  +  \frac{\kappa_1}{2} (e^{4\beta J}-1)\mathcal{D}(A_i(4J))[\rho_S]+
     \frac{\kappa_1}{2} (e^{8\beta J}-1)\mathcal{D}(A_i(8J))[\rho_S] \nonumber \\
    &=  \frac{\kappa_1}{2} \mathcal{D}(a_i+a_i^{\dag})[\rho_S] +   \frac{\kappa_1}{2} (e^{4\beta J}-1)\sum_{\sigma(3)}\mathcal{D}((a_i+a^{\dag}_i) P_{\sigma(3)})[\rho_S]
    +
     \frac{\kappa_1}{2} (e^{8\beta J}-1)\sum_{\sigma(4)}\mathcal{D}((a_i+a^{\dag}_i) P_{\sigma(4)})[\rho_S].
\end{align}
We can now choose $\beta = \frac{1}{8J}\ln\left[\frac{\kappa_{nn}+\kappa_1}{\kappa_1}\right]$ for some $\kappa_{nn}$. Then we have
\begin{equation}
    \frac{\kappa_1}{2}\mathcal{D}(a_i+a_i^{\dag})[\rho_S] +  \frac{\tilde{\kappa}_{nn}}{2}\sum_{\sigma(3)}\mathcal{D}((a_i+a^{\dag}_i) P_{\sigma(3)})[\rho_S]
    +
   \frac{\kappa_{nn}}{2}\sum_{\sigma(4)}\mathcal{D}((a_i+a^{\dag}_i) P_{\sigma(4)})[\rho_S],
\end{equation}
where $\tilde{\kappa}_{nn} = \sqrt{ \kappa_1 \kappa_{nn} + \kappa_1^2} -\kappa_1 $. This is almost the same as the dissipators defined in Eq.\ (\ref{eq:nn}) of the main text. The first difference is that all jumps here are proportional to $a_i + a_i^\dagger$, while in the main text all jumps are proportional to just $a_i$. (We shall address this in the next paragraph.) The second difference is that here $\kappa_1$ came solely from the bath that we designed ourselves, representing the fluctuations associated with a non-zero-temperature bath that disorders the state. We note that single-photon loss processes can also occur due to spontaneous emission for a single cavity, which arises due to    processes that lower the energy within a single cavity (not included in the analysis above for simplicity). 
%is physically a different source of noise that flips the state's parity. 
Taking such terms into account amounts to shifting up the effective temperature of the model. However, as long as the composite system sits in the low-temperature part of the phase diagram (i.e.~if the ``zero-temperature'' part of the bath is significantly stronger than the terms  that generate bit flips),  the system will sit in the part of the phase diagram that hosts a quantum memory.

To recover the photonic-Ising dissipators, we take into account the two-photon process on each photonic-Ising site. The two-photon process effectively constrains the bosonic Hilbert space on each site to a two-dimensional manifold, making the thermal equilibrium of $H_S$ well-defined. Without the two-photon process, $H_S$ is infinitely degenerate even for a finite photonic-Ising lattice, which leads to an ill-defined steady-state manifold for the Lindbladian even when detailed balance is enforced.

To simplify the analysis, we assume %that the two-photon drive is large so we have quantum Zeno type dynamics
that $N = |\alpha|^2 \rightarrow \infty$ and $\kappa_2$ is nonzero, in which case
 the system is at all times close to the ideal manifold spanned by the two coherent states $\ket{\pm \alpha} = \ket{\pm e^{- i \pi/4} \sqrt{N}}$  with %.  We now take the limit $N \rightarrow \infty$, in which case
 $\langle \alpha|{- \alpha}\rangle = 0$. So, within this two-dimensional Hilbert space, %We note that for coherent states $\ket{e^{i\theta}\sqrt{N}}$, we may approximate $a^{\dag}\sim e^{-2i\theta}a$ at $N\to\infty$. For $\theta = \pi/4$ as in the main text, we have 
both $a$ and $a^\dagger$ have vanishing diagonal matrix elements, while their off-diagonal matrix elements are off by only a phase, so we can write  
$a+a^{\dag} \rightarrow (1+i)a$. This finally yields
\begin{equation}
     \kappa_1\mathcal{D}(a_i)[\rho_S] +  \tilde{\kappa}_{nn}\sum_{\sigma(3)}\mathcal{D}(a_i P_{\sigma(3)})[\rho_S]
    +
   \kappa_{nn}\sum_{\sigma(4)}\mathcal{D}(a_i P_{\sigma(4)})[\rho_S],
\end{equation}
which are the photonic-Ising dissipators [Eq.\ (\ref{eq:nn}) in the main text] and the single-photon loss noise at site $i$. 

Note that, while we made many assumptions to arrive at the precise dissipators from Eq.\ (\ref{eq:nn}) in the main text, it is likely that nearly any low-temperature thermal Markovian bath will result in a protected quantum memory. 

% It is worth noting that we have made use of several approximations during the derivation. The extent to which these assumptions are met is to be further verified against specific experimental settings. In particular, we assumed the coupling $H_{SB}=\sum_i (a_i+a^{\dag}_i)\otimes B_i $ does not cause strong coupling between the bath and the system, and the resulting disturbance to the bath relaxes quickly. One has to be cautious when the two-photon drive pumps a large number of photons into the cavity. In this case $|\langle a\rangle|\sim \sqrt{N}$, where $N$ is the average photon number in the cavity. The coupling $H_{SB}$ thus gets amplified as $N$ grows and the assumption might eventually break down. However, we could still attain a robust memory that lives long enough for practical purposes before this breakdown happens. It is also interesting to investigate to whether we can maintain the robust quantum memory beyond these assumptions.

\section{5.~Engineering an Ising interaction between cavity modes}

\begin{figure}[h]
    \centering
    \includegraphics[scale=0.3]{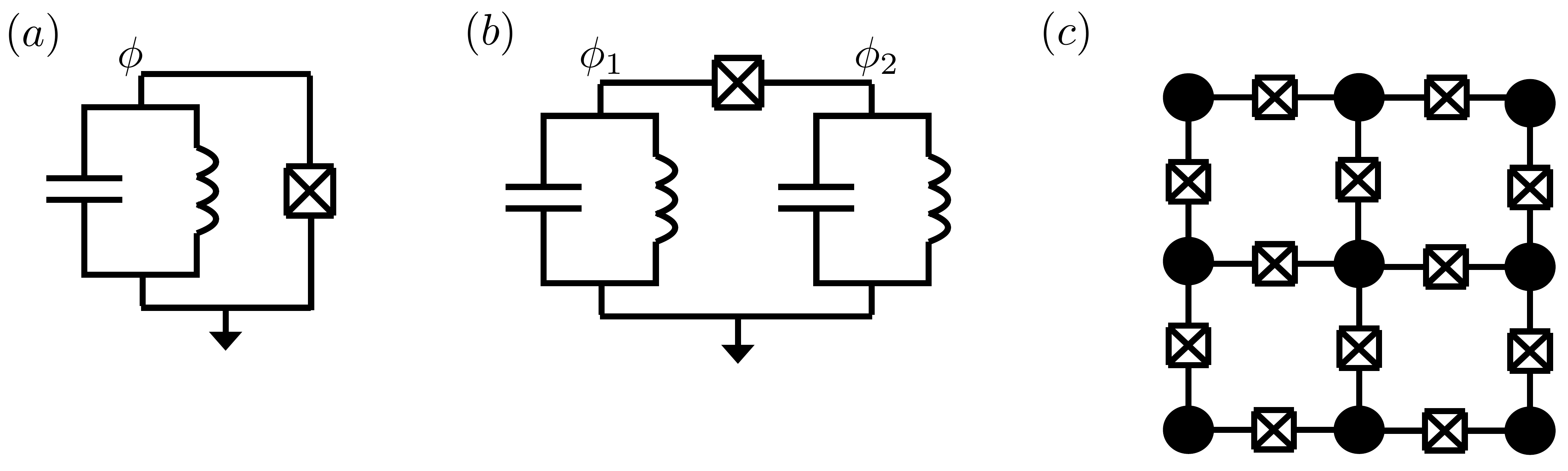}
    \caption{ (a) An $LC$ oscillator (representing a cavity) connected to a Josephson junction. The junction generates a Hamiltonian term that is proportional to the photon parity in the cavity. (b) Two $LC$ oscillators connected via a Josephson junction. The junction generates an Ising-like parity-parity interaction between the oscillators in a certain parameter regime. (c) Schematic layout for a 2D quantum memory: Black circles represent driven-dissipative resonator cavities (or $LC$ oscillators). Each resonator is coupled to its neighbor via a Josephson junction which generates an Ising-like interaction (in the appropriate limit). }
    \label{fig:sm:circuit}
\end{figure}

We showed in the last section that the dissipators described in Eq.~\eqref{eq:nn} most naturally appear  as thermal dissipators corresponding to  processes that raise and lower the energy of the Ising-like Hamiltonian $H = - J\sum_{\langle i j \rangle} Q_i Q_j$, where $Q_i = e^{i \pi a_i^\dagger a_i}$.  The most straightforward  approach to achieve a passively protected memory is therefore to engineer an Ising-like interaction between nearest-neighbor cat qubits. At low temperatures, the coupling of the system to its thermal environment can lead to thermal processes that drive the system to a ferromagnetic state, thus protecting the system against bit flips.  

Previous studies have described how to achieve a parity-parity interaction term between neighboring cat qubits \cite{cohen2017, cohenthesis}. Here we first review the steps needed to achieve a single-cavity Hamiltonian proportional to parity, closely following Ref.~\cite{cohen2017}. We then discuss  the generalization to a parity-parity interaction which follows in a very similar manner.

 Consider an LC oscillator (representing a cavity mode) connected to a Josephson junction, as shown in Fig.\ \ref{fig:sm:circuit}(a), in the presence of a two-photon drive on the cavity (not shown).  The Hamiltonian of the system reads
\begin{equation}
H  = \frac{\hat q^2}{2C} + \frac{\hat \phi^2}{2L} - E_J \cos \left(  \hat{\phi} / \Phi_0 \right) + \lambda \left( \hat{a}^2 e^{i \omega_d t} + (\hat{a}^\dagger)^2 e^{- i \omega_d t} \right),
\end{equation}
where $\hat q $ is the charge of the capcitor $C$, $\hat\phi$ is the flux through the inductor $L$, and $E_J$ is the Josephson energy,   $\Phi_0 = \hbar / (2e)$ is the flux quantum, $\hat{a}$ annihilates an excitation of the cavity, $\lambda$ is the drive strength, and $\omega_d$ is the drive frequency. We note that the flux through the inductor is equal in magnitude to the flux through the Josephson junction since Kirchoff's law relates the voltage drop across the two elements at any given time. We can rewrite the Hamiltonian in terms of $a, a^\dagger$:
\begin{equation}\label{sm:eq:h_cavity}
H  = \hbar \omega (a^\dagger a + 1/2)- E_J \cos{[ x (a + a^\dagger)]} + \lambda \left( a^2 e^{i \omega_d t} + (a^\dagger)^2 e^{- i \omega_d t} \right) ,
\end{equation}
where $\phi = \sqrt{\hbar Z / 2} (a+a^\dagger), q = (1/i)\sqrt{\hbar/(2Z)}(a-a^\dagger),   \omega = 1/\sqrt{ L C}, Z =\sqrt{L/C}, x = \Phi_0^{-1} \sqrt{\hbar Z/2}$ and we drop the hats on operators henceforth. 

% $\omega$ is the frequency of the cavity, $z = 2 \sqrt{\pi  Z / R_K }$, $R_K = h / e^2$ is the resistance quantum, and $Z = \sqrt{L/C}$ is the impedance. 

Going to the interaction picture with respect to $\hbar \omega (a^\dagger a + 1/2)$, the  Hamiltonian takes the form
\begin{equation}
H_{int}= -  E_J \cos{[x (a e^{-i \omega t} + a^\dagger  e^{+i \omega t})]} + \lambda (a^2 + (a^\dagger)^2) =  - \frac{E_J}{2}   (D[\beta(t)] + D[- \beta(t)] ) + \lambda (a^2 + (a^\dagger)^2),
\end{equation}
where we have assumed an on-resonant drive, $\omega_d = 2 \omega$, and  defined the displacement operator $D[\beta(t)] = e^{\beta(t) a^\dagger - \beta^*(t) a}$, with $\beta(t) =  i x e^{- i \omega t}$. We apply the rotating-wave approximation to remove all time dependence in the Hamiltonian, which is valid in the limit: $\omega \gg E_J$:
\begin{equation}\label{sm:eq:rwa}
    H_{rw} = - E_J e^{-x^2 / 2} \sum_n \left( L_n(x^2) |n \rangle \langle n | \right) + \lambda (a^2 + (a^\dagger)^2),
\end{equation}
where $L_n$ is the Laguerre polynomial of order $n$. 

The two-photon drive (along with engineered two-photon loss) ensures that the system is approximately confined to a two-dimensional manifold spanned by the cat states $|\alpha_e \rangle$ and $|\alpha_o \rangle$.
%The role of the two-photon drive serves 
%In the presence of two-photon drive and two-photon loss, we can assume that the system is approximately %(to lowest order) 
%confined to a two-dimensional manifold spanned by the cat states $|\alpha_e \rangle$ and $|\alpha_o \rangle$. 
In this subspace, the Hamiltonian is diagonal (since $H_{rw}$ is diagonal in the Fock basis). Further, if we set $x = 2|\alpha|$, then the Hamiltonian is exponentially close to the parity operator:
\begin{align}
    H_{rw}^{\text{cat}}  &= - \frac{\hbar \Omega}{2} (|\alpha_e \rangle \langle \alpha_e | - |\alpha_o \rangle \langle \alpha_o |) + O(E_J e^{- |\alpha|^2 / 2}) \\
    &= - \left( \frac{\hbar \Omega}{2} \right) Q + O(E_J e^{- |\alpha|^2 / 2}), \qquad Q = e^{i \pi a^\dagger a}, \label{eq:parity_H}
\end{align}
where $\Omega  = E_J / (\hbar \sqrt{2\pi|\alpha|^2})$.  The exponentially small term is proportional to the identity in this subspace, which should not affect the dynamics. 

Let us briefly summarize the physical conditions required to achieve a parity Hamiltonian of the form $H \sim Q$ in \eqref{eq:parity_H} above. In the large $|\alpha|^2 \equiv N$ limit, we require that $E_J \sim \sqrt{N}$ such that the coefficient $\Omega$ does not tend to zero at large $N$. In order for the rotating wave approximation to be valid, we required that $\omega \gg E_J$. This implies that $\omega \sim \sqrt{N}$, which can be achieved with a small capacitance: $\omega \sim 1/\sqrt{L C}$ if $C \sim N^{-1}$. Finally, we required $x \sim \sqrt{L/C} \sim \sqrt{N}$. This is again satisfied with the small capacitance condition: $C \sim N^{-1}$. 

%First, we asumed the rotating wave approximation

So far, our discussion has focused on achieving a parity Hamiltonian for a single cavity. A very similar setup will result in a  parity-parity interaction between neighboring cavity modes \cite{cohen2017}. We briefly describe how this can be done. 

Consider two driven LC oscillators which are connected to a Josephson junction, as shown in Fig.\ \ref{fig:sm:circuit}(b). 
(We again assume two-photon drives on each cavity as before, but neglect to include them in the Hamiltonian since they only serve the purpose of confining the state of the system to the cat state subspace, as described in the previous paragraphs.) 
The Hamiltonian for the system reads
\begin{equation}
    H=   \left( \frac{q_1^2}{2 C_1} +  \frac{\phi_1^2}{2 L_1} \right)   +  \left(\frac{q_2^2}{2 C_2} +  \frac{\phi_2^2}{2 L_2} \right)   - E_J \cos ( \phi_2 + \phi_1) ,
\end{equation}
where $\phi_{1/2}$ are the node fluxes defined in Fig.\ \ref{fig:sm:circuit}(b). Moving to the rotating frame of the cavity Hamiltonians leads to the interaction Hamiltonian
\begin{equation}
    H_{int} = -E_J \cos \left( (a_2 e^{-i \omega_2 t} + a_2^\dagger e^{i \omega_2 t} ) + (a_1 e^{-i \omega_1 t} + a_1^\dagger e^{i \omega_1 t} ) ) \right),
\end{equation}
where $a_{1/2}$ are the annihilation operators associated with the two cavities, and $\omega_{1/2}$ are the frequencies.  Applying the rotating-wave approximation leads to
\begin{equation}
    H_{rw} =  -E_J e^{ -(x_1^2 + x_2^2)/2} \sum_{n_1, n_2} L_{n_1}(x_1^2) L_{n_2}(x_2^2) |n_1, n_2 \rangle \langle n_1, n_2 | .
\end{equation}
Note that we also require that the frequencies of the cavities should be incommensurate in order for the terms above to be the only ones that are time independent, i.e.~$l_1 \omega_1 \neq l_2 \omega_2, \forall l_1, l_2 \in \mathbb{Z}$. 
Specializing to the two-dimensional cat state manifold leads to an interaction of the form
%We can achieve a similar parity-parity interaction between two cavities by coupling two LC oscillators in series to a Josephson junction \cite{cohen2017}. Repeating the steps above leads to an interaction of the form
\begin{equation}
    H_{rw}^{\text{cat}} = - \left( \frac{    \Omega_1 \Omega_2}{4 E_J} \right) Q_1 Q_2,  \qquad Q_{1/2} = e^{i \pi a_{1/2}^\dagger a_{1/2}}, \label{eq:parity_H2}
\end{equation}
where $\Omega_{1/2}  = E_J / (  \sqrt{2\pi|\alpha_{1/2}|^2})$  set the energy scale of the coupling between the two cavities $1,2$.

 Again, let us briefly summarize the physical conditions required to achieve the parity-parity  Hamiltonian of the form $H \sim Q_1 Q_2$ in Eq.\ \eqref{eq:parity_H2} above. In the large $|\alpha|^2 \equiv N$ limit, we require that $E_J \sim N$ such that the coefficient $\sim \Omega_1 \Omega_2/E_J$ does not tend to zero at large $N$. In order for the rotating wave approximation to be valid, we required that $\omega_{1/2} \gg E_J$ and $|\omega_1-\omega_2|\gg E_J$. This implies that $\omega \sim N$. We also require that  $x_{1/2} \sim \sqrt{L_{1/2}/C_{1/2}} \sim \sqrt{N}$. Both of these conditions can be achieved for $C_{1/2} \sim N^{-3/2}, L_{1/2} \sim N^{-1/2}$.
 
 A schematic layout for a 2D quantum memory is provided in Fig.\ \ref{fig:sm:circuit}(c): Each black dot represents a driven-dissipative resonator, connected to its neighbor via a Josephson junction that results in an Ising interaction. The advantage of the passive approach is the lack of ancilla qubits and of precise pulse signals that are typically needed to make the measurements required for active error correction. The passive approach also avoids the need for classical communication with a decoder.

 The scheme we have just described to achieve a parity-parity interaction requires large tunability of the frequency and impedance of a cavity, and of the Josephson tunneling coefficient of the junction. These conditions may prove to be challenging to engineer in practice, and it is still unclear whether such a protocol is the simplest way to arrive at the effective model described in the main text. Any circuit QED proposal to arrive at our model should involve nonlinear devices, since the desired parity-parity interaction is highly nonlinear.  In this section,  we have explored a conceptually simple route to achieve this nonlinearity via a single Josephson junction. It is conceivable that  adding other devices would relax the conditions required to achieve a substantial  Ising interaction, necessary for an exponentially suppressed logical bit flip rate. We leave such investigations to future work.

\section{6.~Digital autonomous photonic-Ising local decoder}

In this section, we describe a digital autonomous approach for realizing a stochastic local error decoder inspired by the photonic-Ising dissipators defined in the main text. This procedure involves implementing a sequence of fault-tolerant local gates to correct the errors without the need of measurements. The step is then iterated over time on the entire system as fast as possible. Although this approach is different from directly realizing the microscopic Lindbladian (which could be done by dividing the error-correcting step infinitesimally~\cite{Weimer2010, Barreiro:2011}), this digital approach is potentially easier to realize experimentally, and we expect that it provides the same dynamical protection of the quantum memory. Note that the protocol requires the rate of the digital steps to scale linearly with the average photon number. 
%We will describe an analog approach to directly realize the photonic-Ising model in the next section.

It suffices to consider the implementation of the local decoder at a single time step. Consider a square lattice of photonic cavities. Over each lattice cavity, we place an ancillary cavity (initialized in $\ket{\alpha_o}$). 

We describe the implementation at a single site and the generalization to the whole lattice follows straightforwardly.
First, for a given site, we perform an encoding unitary $U$. %that acts on the chosen lattice cavity, the four neighboring lattice cavities, and the corresponding ancillary cavity. %\mathcal{H}_{\text{lattice}}\otimes\mathcal{H}_{\text{ancilla}}$. 
Depending on the state of the chosen lattice cavity and its four neighbors, $U$ changes the state of the ancillary cavity via
\begin{equation}
    U = P\otimes (\ketbra{\alpha_e}{\alpha_o} + \ketbra{\alpha_o}{\alpha_e})+ P^{\perp}\otimes (\ketbra{\alpha_o}{\alpha_o}+\ketbra{\alpha_e}{\alpha_e}), \label{sm:eq:encoding}
\end{equation}
where $P$ projects on a local configuration of domain walls (a specific example was considered in the main text, where we project onto a configuration with 3 or 4 misaligned neighboring lattice cavities) and $P^{\perp}$ is the orthogonal subspace projector.
We define the unitary $U$ such that it changes the ancillary cavity from $\ket{\alpha_o}$ to $\ket{\alpha_e}$ if a local error is detected; it does nothing otherwise. Note that $U$ can be implemented using the fundamental set of bias-preserving gates in Ref.~\cite{prx2019}, where the two-photon drive and two-photon loss can be kept on thus suppressing the dephasing errors during the gate implementation. 

Second, we apply a CNOT gate (described in Section IV.D in Ref.~\cite{prx2019}) that is controlled by the ancillary cavity and targets the corresponding lattice cavity. (We use the convention that $\ket{\alpha_o}$ is $\ket{0}$ and $\ket{\alpha_e}$ is $\ket{1}$). 

Third, we use a strong dispersive coupling to a transmon to extract the entropy from the ancillary cavity~\cite{leghtas2013} and reset it back to the initial state $\ket{\alpha_o}$. Due to the fault-tolerance of the cavity-cavity gates~\cite{prx2019}, the phase errors stay suppressed when the Ising-type local decoder is implemented autonomously.

The full procedure is achieved by implementing the encoding and the reset across the entire lattice.
To extend the single-site procedure to the entire lattice, we note that the encoding operations $U$ on each site are local around each lattice cavity and they commute across different lattice sites. Therefore, the encoding $U$ can be implemented in parallel across all the sites before a final reset, e.g. by dividing the lattice into bipartite sublattices and operating on the cavities that belong to the same sublattice in parallel. 

%In summary, we have the following steps: 
%\begin{enumerate}
%    \item
%    Randomly pick a lattice cavity and apply $U$ (constructed from the fundamental set of gates in Ref.~\cite{prx2019}). If the ancillary cavity of that site is $\ket{\alpha_e}$,  we need to flip the parity of the central lattice; otherwise, we do not need to do anything.
%   \item 
%    Apply a CNOT (described in Section IV.D in Ref.~\cite{prx2019}) controlled by the ancillary cavity and targeting the corresponding lattice cavity.
%    \item 
%    Incoherently reset the ancillary cavity to the initial odd parity state by strong dispersive coupling to a transmon and a transmon reset (described in Ref.~\cite{leghtas2013}).
%\end{enumerate}

\begin{figure}[t]
    \centering
    \includegraphics{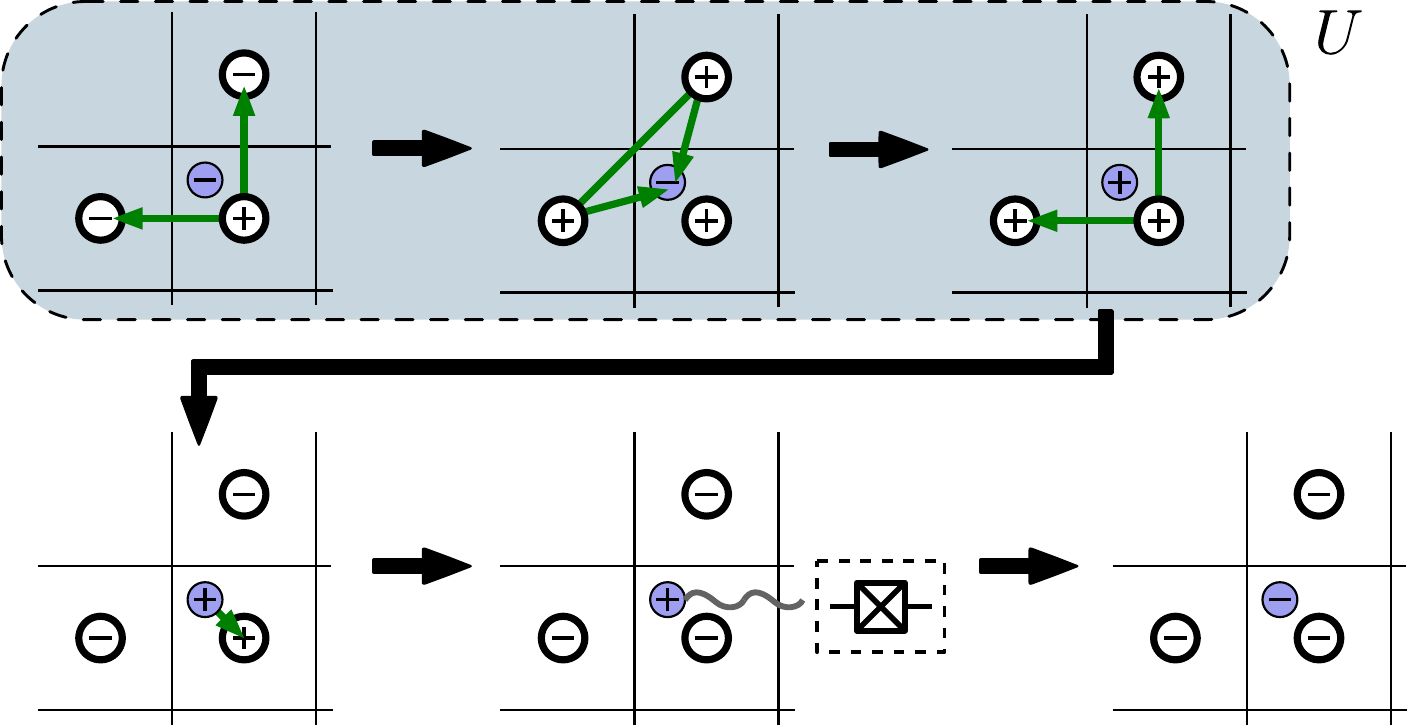}
    \caption{Illustration of the digital autonomous photonic-Ising local decoder when $U$ is implementing Toom's rule. Purple color labels the ancillary cavity. We denote $\ket{\alpha_o}$ by ($-$) and $\ket{\alpha_e}$ by ($+$). The green arrows are CNOT gates, pointing from the control cavity to the target cavity. A connected double green arrow is a Toffoli gate. The procedure consists of the following steps: (i) two CNOTs from the central cavity to its neighbors; (ii) a Toffoli gate from the neighbors to the ancillary cavity; (iii) two CNOTs from the central cavity to its neighbors; (iv) a CNOT from the ancillary cavity to the central cavity; (v) reset the ancillary cavity via coupling to a transmon.}
    \label{fig:sm:autonomous_examle}
\end{figure}

For concreteness, let us consider an example where $U$ flips the ancillary cavity if it identifies a corner formed by the domain walls, i.e.~Toom's rule. (Digitally, this is simpler than the majority rule described in the main text, but one could implemenent the majority rule approach as well.) Then $U$ for a chosen orientation can be implemented by the following gate sequence
\begin{enumerate}
    \item Apply two CNOTs from center lattice cavity to its neighboring cavities on the left and on the top.
    \item Apply a Toffoli gate (described in Section IV.E in Ref.~\cite{prx2019}) controlled by the two neighboring cavities and targeting the ancillary cavity.
    \item Repeat step 1 to invert the two CNOTs applied.
\end{enumerate}
The digital procedure for the autonomous implementation of Toom's rule on a particular configuration is schematically depicted in Fig.~\ref{fig:sm:autonomous_examle}.

The autonomous approach above can be easily turned into an active error correction protocol: instead of applying a CNOT from the ancillary cavity to the central cavity and then resetting the ancillary cavity, we can measure the ancillary cavity and flip the parity of the central cavity if the measurement result is ($+$). %by replacing the step of incoherent reset with measurement and feedback control. 
Alternatively, when implementing active error correction, we can place a syndrome cavity between each pair of neighboring cavities on the lattice. By storing in the syndrome cavity the information regarding the presence of a domain wall, we can implement a local decoder based on the rules defined by $U$. This then becomes a 2D version of a repetition cat code. Again, all the steps can be achieved with dephasing errors exponentially suppressed.  In contrast with the non-local processing of syndrome information required by the 1D repetition cat code in Ref.~\cite{prx2019}, the 2D code allows for a stochastic local decoding procedure. As mentioned at the beginning of the section, to achieve an exponentially long memory time, both the autonomous and the active error correction approaches require scaling the rate of the digital step linearly with the average number of cavity photons $N$ because bit-flip error rate scales with $N$.

\end{document}